\documentclass[journal]{IEEEtran}
\usepackage{subfigure}
\usepackage{graphicx}
\usepackage{xcolor}
\usepackage{amsmath}
\usepackage{mathtools}
\usepackage{amssymb}
\usepackage{verbatim} % 用来使用多行注释：\begin{comment}, \end{comment} 
\usepackage{diagbox}
\usepackage{hyperref} % 把链接弄成可以点击的
\usepackage{multirow} % table row合并？
\usepackage{algorithmic}
\usepackage{textcomp}
\usepackage{threeparttable} % 给table加下面的注释
\usepackage[normalem]{ulem} % performance table用加下划线？
\useunder{\uline}{\ul}{} % performance table用加下划线？

\ifCLASSINFOpdf
\else
\fi
\begin{document}

\title{Perceptual Quality Assessment of UGC Gaming Videos}
 \author{Xiangxu Yu, Zhengzhong Tu, Neil Birkbeck, Yilin Wang, Balu Adsumilli and Alan C. Bovik
 \thanks{X. Yu, Z. Tu and A. C. Bovik are with the Department of Electrical and Computer Engineering, University of Texas at Austin, Austin, USA (e-mail: yuxiangxu@utexas.edu; zhengzhong.tu@utexas.edu; bovik@ece.utexas.edu). Neil Birkbeck, Yilin Wang and Balu Adsumilli are with YouTube Media Algorithms Team, Google Inc (e-mail: birkbeck@google.com; yilin@google.com; badsumilli@google.com).}
 }


\markboth{}%
{}

\maketitle

\begin{abstract}
In recent years, with the vigorous development of the video game industry, the proportion of gaming videos on major video websites like YouTube has dramatically increased. 
However, relatively little research has been done on the automatic quality prediction of gaming videos, especially on those that fall in the category of ``User-Generated-Content'' (UGC). 
Since current leading general-purpose Video Quality Assessment (VQA) models do not perform well on this type of gaming videos, we have created a new VQA model specifically designed to succeed on UGC gaming videos, which we call the Gaming Video Quality Predictor (GAME-VQP). 
GAME-VQP successfully predicts the unique statistical characteristics of gaming videos by drawing upon features designed under modified natural scene statistics models, combined with gaming specific features learned by a Convolution Neural Network. 
We study the performance of GAME-VQP on a very recent large UGC gaming video database called LIVE-YT-Gaming, and find that it both outperforms other mainstream general VQA models as well as VQA models specifically designed for gaming videos. 
The new model will be made public after paper being accepted. 
\end{abstract}


% \fi
\IEEEpeerreviewmaketitle

\section{Introduction}

People’s daily lives are filled enormous amounts of many types of digital information, and increasingly, online visual content. 
From social images on Facebook to shared videos on YouTube, digital visual content is found everywhere. 
We will find it convenient to identify two broad categories of mainstream online video content. 
First, streaming television and cinema providers, like Netflix, Hulu, and Amazon Prime, mainly provide high-quality, expertly captured and produced Professional-Generated-Content (PGC) videos, that are created using high-end production-grade equipment. 
Conversely, social video sites like YouTube, TikTok, and Facebook, ingest, process, and stream User-Generated-Content (UGC) that have been mostly captured and uploaded by untrained users having uncertain skills and inexpensive photographic devices. 
Of course, these two categories are by no means entirely exclusive, as social sites may provide high-quality content, and television/cinema streamers may offer older or low-quality products. 

Because of the explosive success of the video gaming industry in recent years, gaming videos now represent a significant and growing fraction of Internet traffic. 
Gaming video streaming is a rapidly growing market and, includes, but is not limited to, game live broadcasts, online games, cloud gaming services, and recorded video gameplays uploaded and shared by users. 
Among these are real-time streaming services, such as Steam online games and Facebook cloud games. 
Other examples are passive gaming videos shared by users, such as gameplay live broadcasts on Twitch, and UGC gaming videos shared on YouTube. 

As with providers of other types of content, gaming video streamers are deeply concerned with being able to provide content having the highest possible visual quality to millions of viewers. 
To accomplish this, they monitor and control compression and other processing steps using perceptually-designed video quality assessment (VQA) tools. 
However, existing popular VQA algorithms may not produce accurate quality predictions on gaming video content, which may not obey statistical assumptions made on naturalistic photographic content. 
Indeed, we show that this is the case, motivating out attempts to produce improved gaming VQA models and algorithms. 
VQA algorithms are generally classified into two categories based on whether they make use of an available high-quality reference videos. 
Accordingly, we will refer to these as Reference models, which include the popular models SSIM \cite{wang2004image} and VMAF \cite{li2016toward}. 
Algorithms that make no use of reference videos, i.e., No-Reference (NR) VQA models, are appropriate in scenarios where no reference videos are available. 
As we will explain in the foregoing, we will be primarily interested in developing tools to predict the quality of UGC gaming videos. 
Since in most applications of interest, UGC videos do not have high quality reference videos available, we will be primarily interested in the development of NR VQA models for UGC gaming videos. 

Gaming videos have gained significant attention in recent years, and the development of VQA algorithms for gaming is becoming a research topic of interest. 
Recent efforts have included the creation of gaming video databases and gaming video quality prediction models. 
Thus far, however, this research has been limited to the quality analysis of PGC gaming videos afflicted by single compression distortions, while there has been little effort applied to the UGC gaming VQA problem. 
This is an important omission, since UGC videos, which are typically affected by any of a wide variety of often severe and commonly commingled distortion types, are becoming quite common. 
Towards filling this gap, we have created a new VQA model specifically designed for UGC gaming videos, which we call the Game Video Quality Predictor (GAME-VQP). 
As we will show, we have found that the new model achieves state-of-the-art (SOTA) results on a large, newly built UGC gaming video quality database, while also generalizing well to generic UGC videos. 

The contributions of our work can be summarized as follows: 
\begin{itemize}
  \item We created a new VQA algorithm called GAME-VQP that is designed to analyze UGC gaming videos. Extensive experiments conducted on a large, recently released UGC gaming video subjective quality database show that, as compared against both existing high-performance general-purpose VQA models, as well as special-purpose gaming video VQA algorithms, the new GAME-VQP model delivers superior performance. 
  \item GAME-VQP is designed around a novel fusion method that combines features drawn from perceptually relevant natural scene statistics (NSS) \cite{ruderman1994statistics} models, with deep learning features. This unusual aggregate combination significantly improves the performance of GAME-VQP. In our approach, NSS features and deep learning features are used to train two independent support vector regressors (SVRs), the responses of which are fused by averaging them to produce the final quality prediction scores. 
  \item As part of this study, we observed that the statistical properties of artificially-generated gaming videos tend to systematically differ from those of naturistic photographic videos. 
  \end{itemize}

The rest of the paper is organized as follows. 
We review background and previous work in Section \ref{related_work}, including a brief description of the recently constructed UGC gaming video database. 
We detail the design of the GAME-VQP model in Section \ref{GAME-VQP_model}. 
We show experimental results and discuss them in Section \ref{performance_evaluation}. 
Lastly, we conclude the paper with some additional salient thoughts in Section \ref{conclusion}. 

\section{Related Work}
\label{related_work}

\subsection{General NR VQA Model}

Creating quantitative models that can blindly predict the perceptual quality of videos is a longstanding, challenging problem. 
There have been a variety of models that have achieved varying degrees of performance. 
The widely-used general-purpose NR VQA models BRISQUE \cite{mittal2012no} and NIQE \cite{mittal2013making} have found commercial success. 
They are based on modifications of classical empirical models of NSS. 
These models posit that the bandpass coefficients of high-quality natural images reliably obey certain statistical laws, that are violated when visible distortions are introduced. 
While BRISQUE and NIQE use simple spatial bandpass models, variations have been devised in the gradient-domain \cite{xue2013gradient}, via wavelet transforms \cite{sampat2009complex}, and by modeling multiple perceptual processes \cite{ghadiyaram2017perceptual}. 
The temporal statistics of natural videos have also been studied to capture temporal impairments.
V-BLIINDS \cite{saad2014blind} was the first such model, analyzing the block DCT statistics of frame-differences. 
Li \textit{et al}. computed 3D-DCT transforms of local space-time regions to extract space-time quality-aware features \cite{li2016spatiotemporal}. 
More recently, TLVQM \cite{korhonen2019two} defined explicit distortion, motion-statistics, and aesthetics features, obtaining promising results on UGC videos. 
VIDEVAL \cite{tu2021ugc} is a more recent high-performance NR-VQA model that relies on careful feature selection, achieving SOTA performances on the largest UGC-VQA databases. 

Recent advances in deep learning have driven the creation of several successful CNN-based approaches to NR VQA modeling. 
An early deep NR-VQA model, VSFA \cite{li2019quality} accounts for content-dependency and temporal-memory effects, by training a temporal Gated Recurrent Unit (GRU) and a subjectively-inspired temporal pooling layer on top of pre-trained ResNets. 
RAPIQUE \cite{tu2021rapique} is a recent hybrid model that combines spatio-temporal NSS features drawn from a variety of perceptual model domains with pre-trained semantics-aware deep learning features. 
These features are jointly used to train a head regressor to predict the overall perceptual video quality. 
RAPIQUE uses a large number of features, but nevertheless is exceptionally fast, owing to clever feature re-use and other efficiencies. 
It achieves SOTA performance on several large NR VQA datasets. 
In another innovation, PVQ \cite{ying2020patch} extracts both local and global spatial and temporal features via RoIPool and SoIPool layers, enabling the learning of local-to-global spatio-temporal quality relationships when trained on a large-scale UGC video quality database containing both local and global subjective labels. 

\subsection{Video Games and Gaming Videos}

\subsubsection{Video Game}
Electronic video games generally refer to interactive games that execute on electronic media platforms such as Personal Computers (PCs), mobile devices, and powerful dedicated cloud gaming platforms based on cloud servers. 
The types of currently popular games, which are often available on multiple platforms, can be divided into such categories as PC games, mobile games, console games, handheld games, VR games, cloud games, and more. 
Games can also be classified according to their mode of play, such as adventure games, action games, first-person shooters, real-time strategy games, fighting games, board games, and massive multiplayer online role-playing games. 
Some games are single-player, while others can be simultaneously played by online participants at the same time. 

\subsubsection{Types of Gaming Videos}
While the term `gaming video' can be broadly used, here we will use it to specifically refer to actual gameplay that is recorded, usually by the participant, with the intention of sharing the recording online. 
These recordings are made using equipment with differing degrees of sophistication and cost. 
As before, we can roughly divide these into the categories PGC and UGC. 
PGC gaming videos are those recorded with professional screen recording software, where the game settings were adjusted to high performance that assumes superior hardware and display capability. 
Gaming videos recorded in this way are of extremely high quality, and generally exhibit little, if any visible distortion. 
By contrast, UGC gaming videos are generally recorded by casual users having limited hardware resources. 
Because of this, the game settings are not set very high, and the recording software generally uses the default settings. 
As a result, the recorded videos usually contain mixtures of distortions having varying degrees of apparentness. 
These include distortions of the game data, the delays or stalls caused by hardware limitations, and compression distortions introduced by the screen recording software. 
Some videos are recorded while broadcasting, and since the live broadcast and screen recording software must share the system resources, further visual disruptions and delays may be included in the recorded videos. 
Because of these diverse distortion processes, the quality of UGC gaming videos is different and often much worse than that of PGC gaming videos. 

\subsection{Gaming VQA Models}
Our current VQA research direction that is directed towards video games involves interactive online cloud games. 
These generate images in cloud servers based on real-time interactions with users, then transmit them to users' client devices. 
Another active research direction involves compressed versions of gaming videos that are recorded in advance and streamed to users who simply view them as non-interactive entertainment videos. 
In this category, most existing work has focused on the before vs. after compression quality analysis of PGC gaming videos. 
To date, little research has been conducted on the NR problem of UGC gaming VQA, e.g., of videos uploaded to YouTube, which is the focus of this paper. 

\subsubsection{Gaming VQA Model}
Quality assessment research on gaming videos originated around 2018. 
Early models like NR-GVQM \cite{zadtootaghaj2018nr}, Nofu \cite{goring2019nofu}, and NR-GVSQI \cite{barman2019no} all operated by extracting simple image features that were used to train shallow regression models implemented as SVRs or Random Forests (RF). 
Because the available gaming video databases they used have only limited quantities of subjective scores, each of these three algorithms instead used VMAF scores as proxy `subjective' training labels. 
Since VMAF is an imperfect quality model, this approach is suboptimal. 
The simple quality features used include the `Spatial Information' (SI) or average Sobel gradient magnitude \cite{winkler2012analysis}, and the Temporal Information (TI) (average absolute frame difference \cite{itu910subjective}), and objective features related to blockiness and blur. 
These models do not include features drawn from perceptual science. 
To ameliorate this, BRISQUE and NIQE quality predictions are also used as features for training. 
A recent model called DEMI \cite{zadtootaghaj2020demi} is a deep learning algorithm that focuses on predicting blockiness and blur. 
DEMI first trains a CNN model, then fine-tunes the parameters on an image database, then finally uses an RF model to output quality prediction results. 
Another algorithm called NDNetGaming \cite{utke2020ndnetgaming}, is also based on a CNN architecture, and also uses VMAF scores as groundtruth proxies for training. 
Recently, the authors of \cite{wen2021subjective} proposed an algorithm based on deep learning for the full reference quality assessment of PGC mobile gaming videos that have been subjected to three types of compression artifacts, by training the models on a non-public dataset. 

\subsubsection{Gaming VQA Database}
Other than the new resource introduced here, there are currently four gaming video quality databases, all specifically designed for gaming VQA studies: GamingVideoSET \cite{barman2018objective}, KUGVD \cite{barman2019no}, CGVDS \cite{zadtootaghaj2020quality}, and TGV \cite{wen2021subjective}. 
GamingVideoSET includes 24 reference videos of unique content from 6 different games, along with 576 compressed versions of them using H.264. 
However, only 90 of the videos have associated subjective scores collected from a human study, which is insufficient to develop effective quality prediction models. 
The design of KUGVD is similar to GamingVideoSET, but there are only 6 reference videos recorded from 6 different games, and 144 compressed videos. 
Again, only 90 of those videos have associated subjective scores. 
CGVDS contains 15 different game recordings and 360 compressed videos, all labeled by subjective data. 
As compared with the previous two databases, CGVDS has more subjective data and has compressed videos under different encoding configurations. 
However, the number of games and reference videos included in CGVDS is limited, and they are all PGC content. 
Unlike databases which only contain PC or console games, like the three just described, TGV is a recently created database containing mobile games. 
It has 150 unique contents recorded from 17 different games and compressed versions of them, totaling 1,293 videos. 

All of the original reference videos used in these four databases were recorded using professional recording equipment and software, and with the game settings set to high quality. 
Hence, the reference videos are of pristine quality without significant visible distortions. 
Since the distortions applied to the pristine videos in these four databases are all just different degrees of a single type of compression distortion, these datasets are useful only for building models for PGC VQA of compressed gaming videos. 
Prior to our work, UGC VQA datasets have not been available.

\subsection{LIVE-YouTube Gaming Video Quality Database}
\label{LIVE-YT-Gaming_intro}
\begin{figure}
\centering
\includegraphics[width = 1\columnwidth]{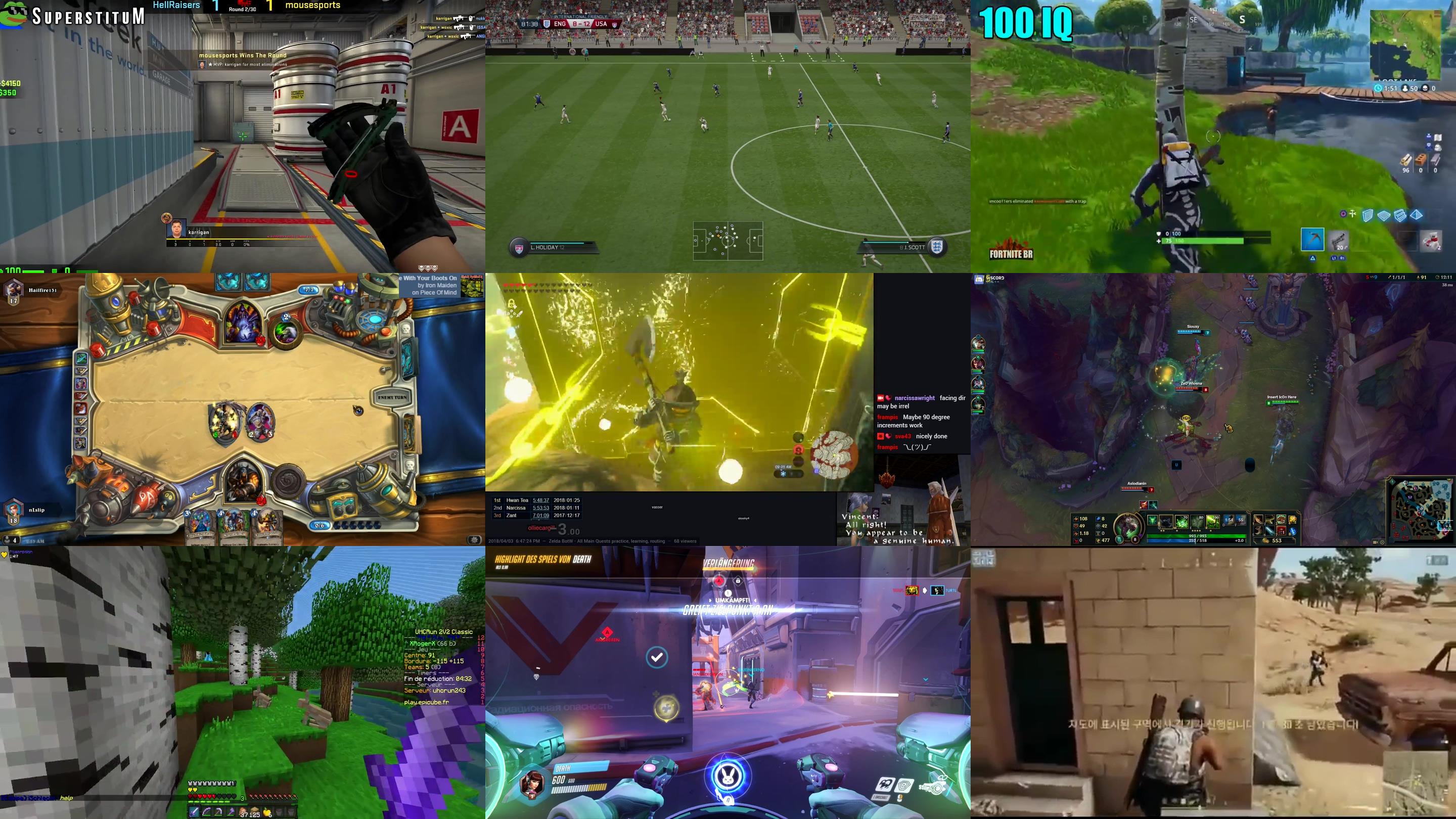}
\caption{Example video frames from the new LIVE-YT-Gaming Database.}
\label{gamevideo_screenshot}
\end{figure}

Recently, a new UGC gaming video quality evaluation dataset was proposed, called the LIVE-YouTube gaming video quality (LIVE-YT-Gaming) Database \cite{yu2022wacv,yu2022gamingvideodatabase}. 
It is the first gaming video database containing only real-world UGC gaming videos and subjective scores recorded on them. 
The LIVE-YT-Gaming database contains a total of 600 unique UGC gaming videos with associated subjective scores, covering different 59 games. 
The videos include 4 resolutions (360p, 480p, 720p, 1080p) and 2 frame rates (30 fps and 60 fps), of videos having durations between 8 and 9 sec. 
Fig. \ref{gamevideo_screenshot} shows example screenshots of some gaming videos in the database. 
An online study was conducted on the videos, which was different from prior crowdsourced VQA studies, since the 61 participants were known reliable naive subjects. 
The study collected a total of 18,600 subjective ratings from 61 subjects, with each video labeled about 30 human viewers. 
This new database substantially addresses the dearth of UGC gaming video quality resources. 
The videos constituting the visual portion of LIVE-YT-Gaming were recorded and uploaded by ordinary, casual users, and are afflicted by mixtures of playback, recording, compression, and other distortions. 
We refer the reader to \cite{yu2022gamingvideodatabase} for substantive details on the content and design of the database and the online study. 
The database has been recently been made publicly available at https://live.ece.utexas.edu/research/LIVE-YT-Gaming/index.html. 

\section{GAME-VQP Model}
\label{GAME-VQP_model}

\begin{figure}
\centering
\includegraphics[width = 1\columnwidth]{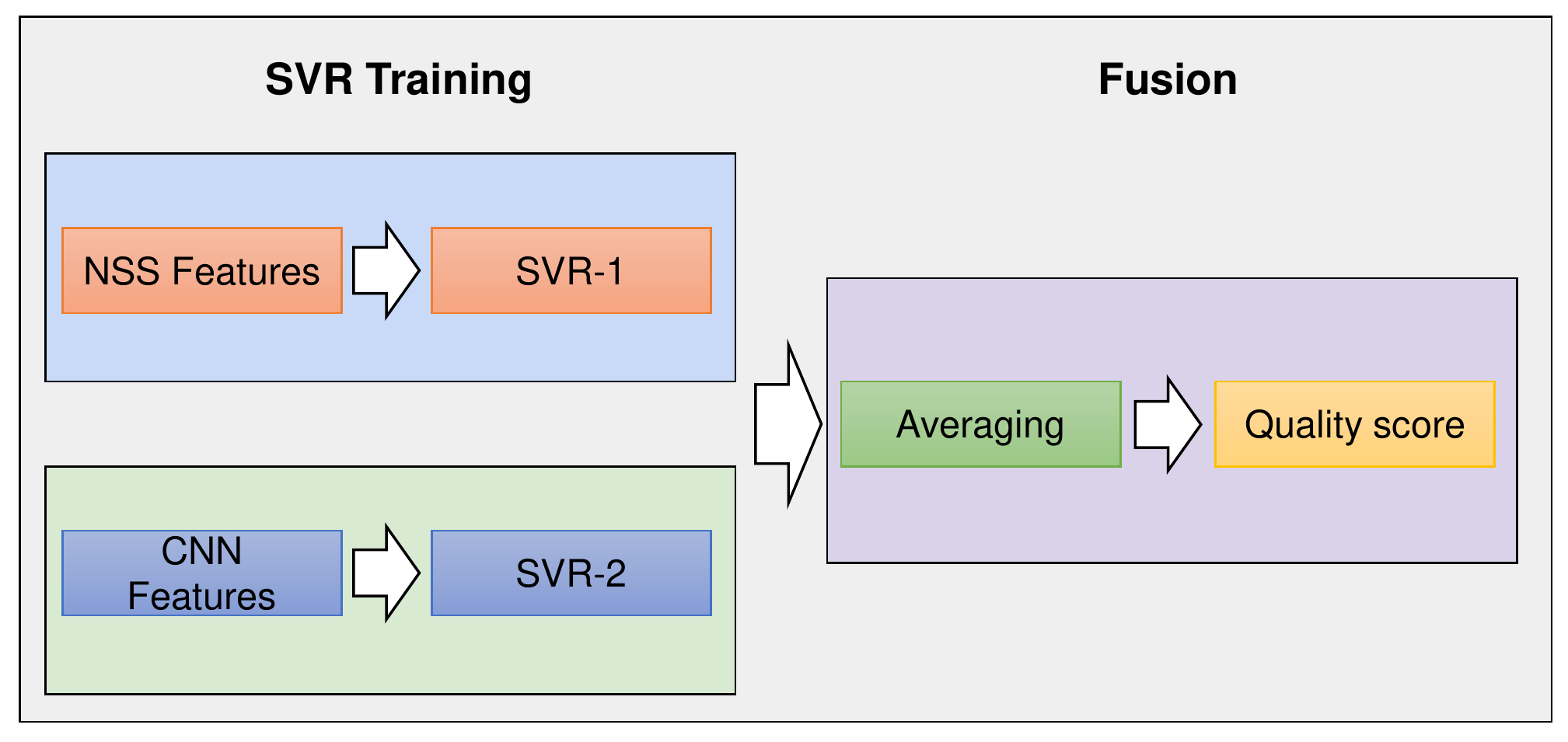}
\caption{Flow chart of the GAME-VQP model.}
\label{model_overview}
\end{figure}

Fig. \ref{model_overview} shows a high-level flow diagram of the GAME-VQP model, consisting of feature extraction, regression modeling, and score fusion modules. 
The feature extraction module includes the computation of both low-level NSS features extracted from training/test videos, as well as high-level features obtained from a pre-trained CNN model.
We trained two separate SVR models on the NSS and CNN features, respectively, reasoning that low-level distortion features and high-level semantic features represent different, independent stages of processing. 
The responses are fused in a simple way to obtain the final video quality predictions, as explained later. 

\subsection{NSS Feature Extraction}

\begin{figure}
\centering
\includegraphics[width = 1\columnwidth]{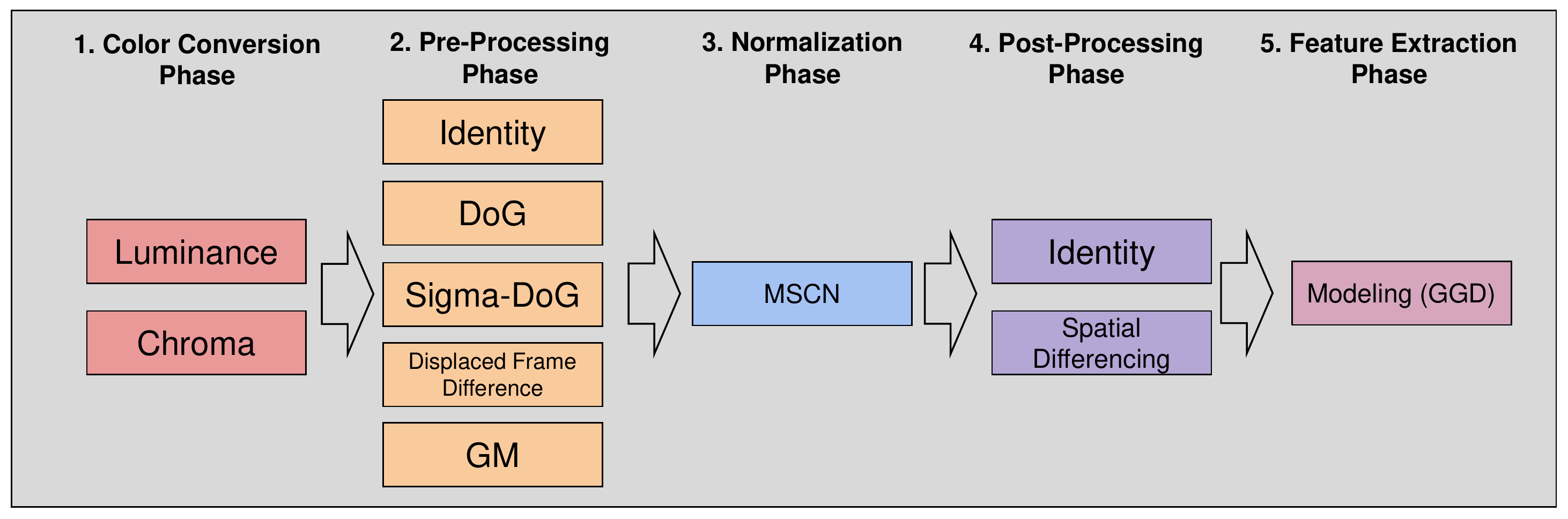}
\caption{Detailed flow chart of NSS feature extraction. }
\label{NSS_overview}
\end{figure}

\begin{figure*}
 \centering
 \footnotesize
 \subfigure[]{
  \label{map_rgb}
  \includegraphics[width=0.23\textwidth]{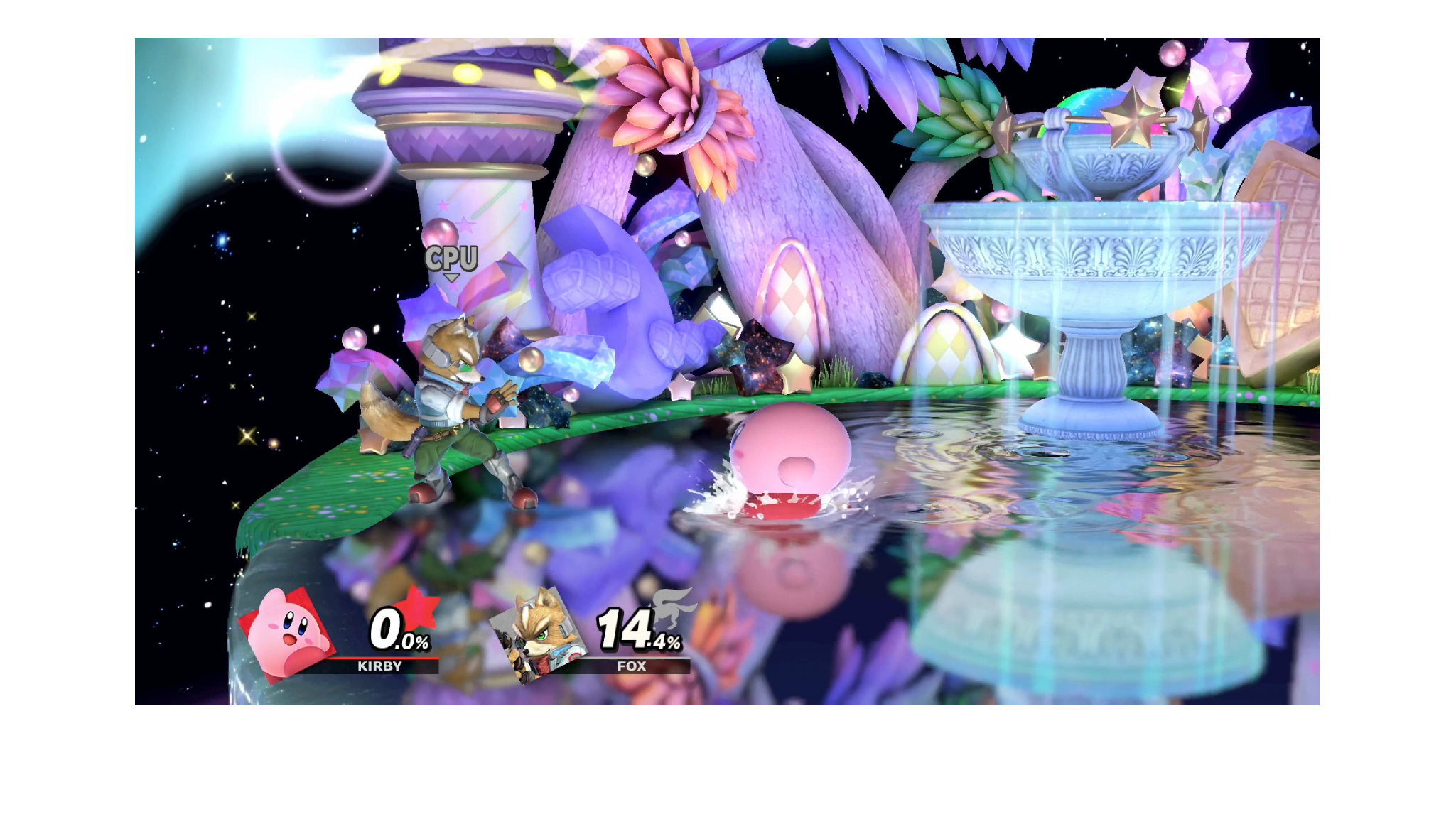}}
 \subfigure[]{
  \label{map_lum} 
  \includegraphics[width=0.23\textwidth]{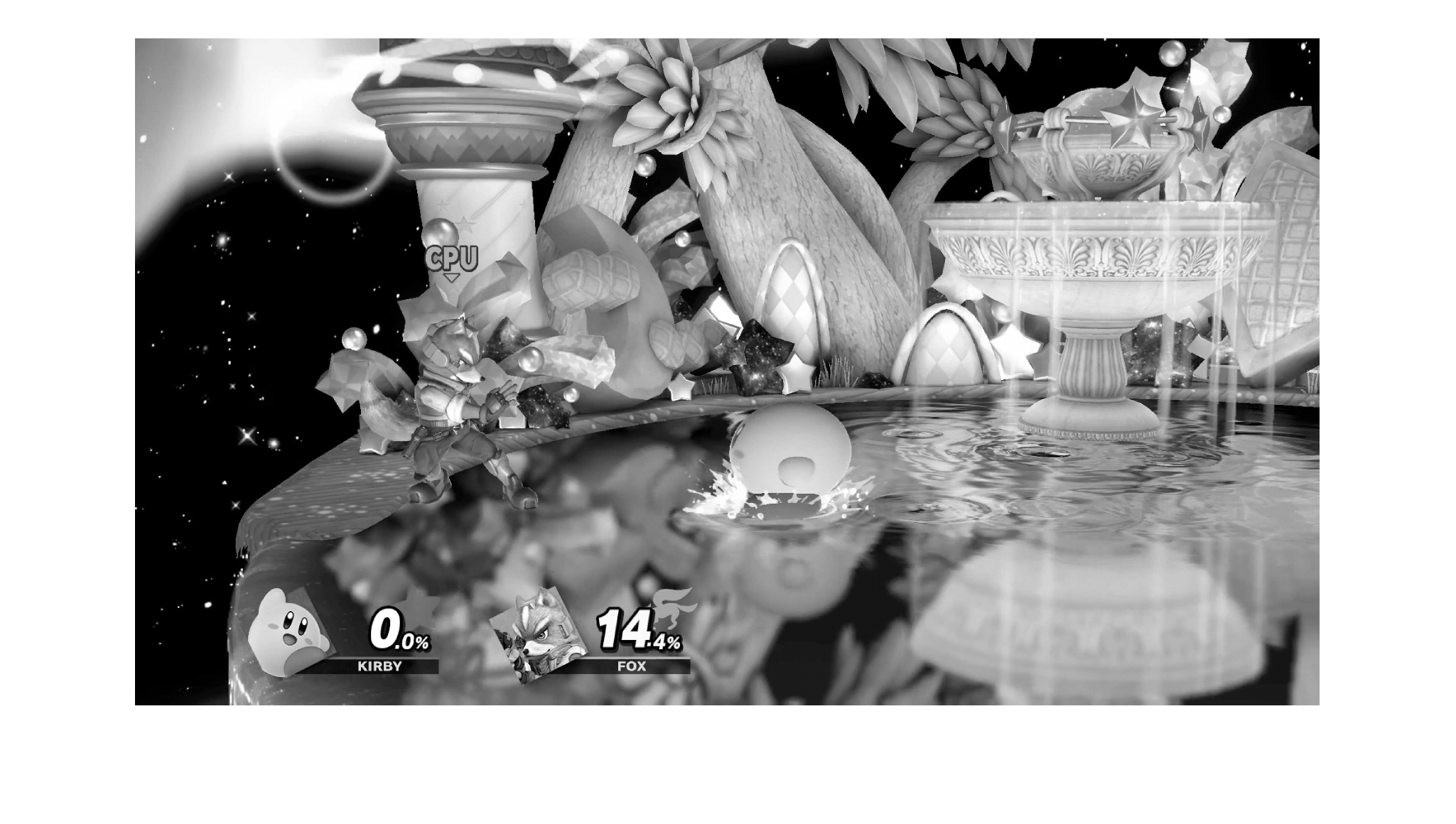}}   
 \subfigure[]{
  \label{map_chroma} 
  \includegraphics[width=0.23\textwidth]{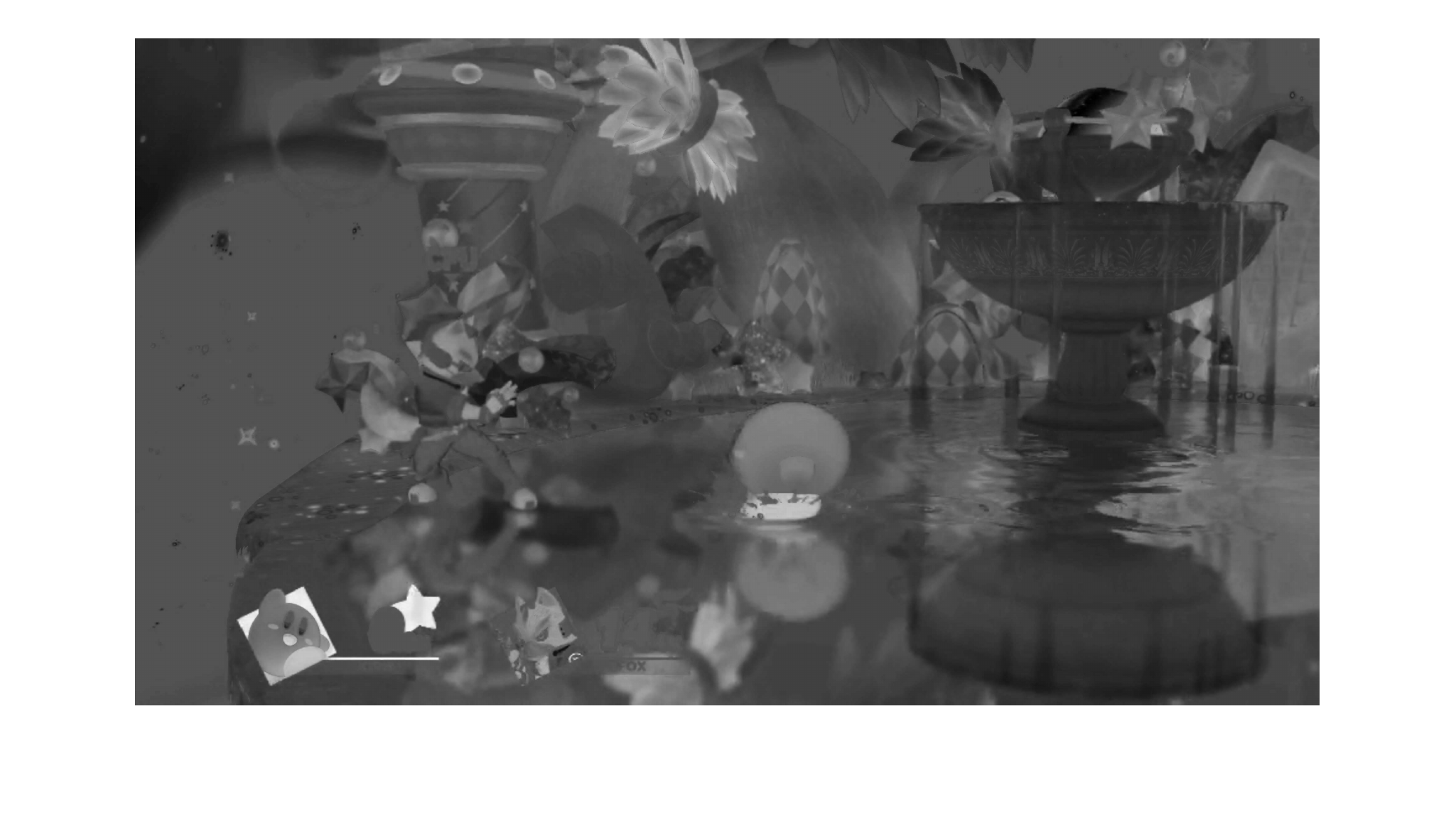}}      
 \subfigure[]{
  \label{map_sigma}
  \includegraphics[width=0.23\textwidth]{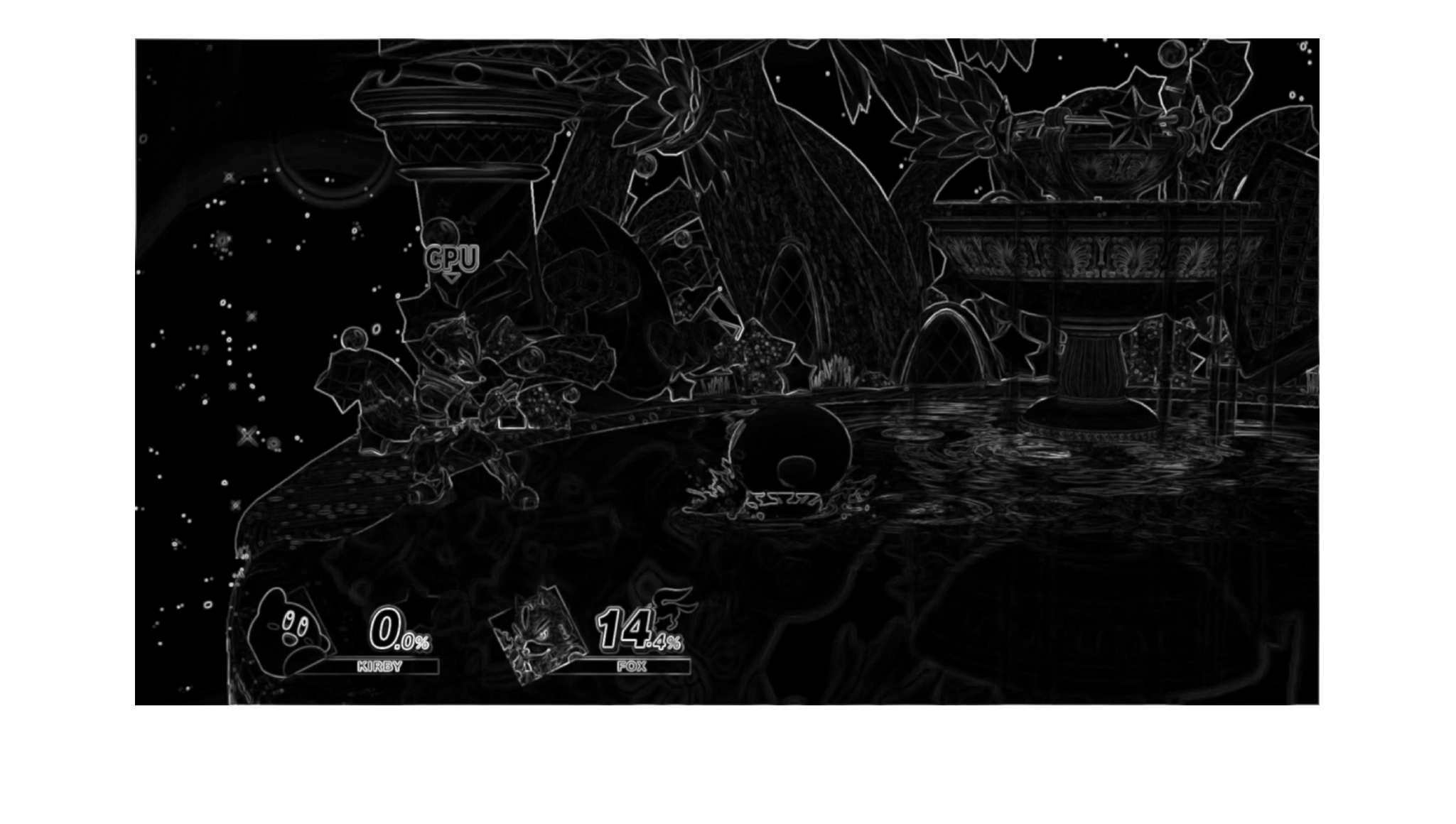}}
 \subfigure[]{
  \label{map_dog} 
  \includegraphics[width=0.23\textwidth]{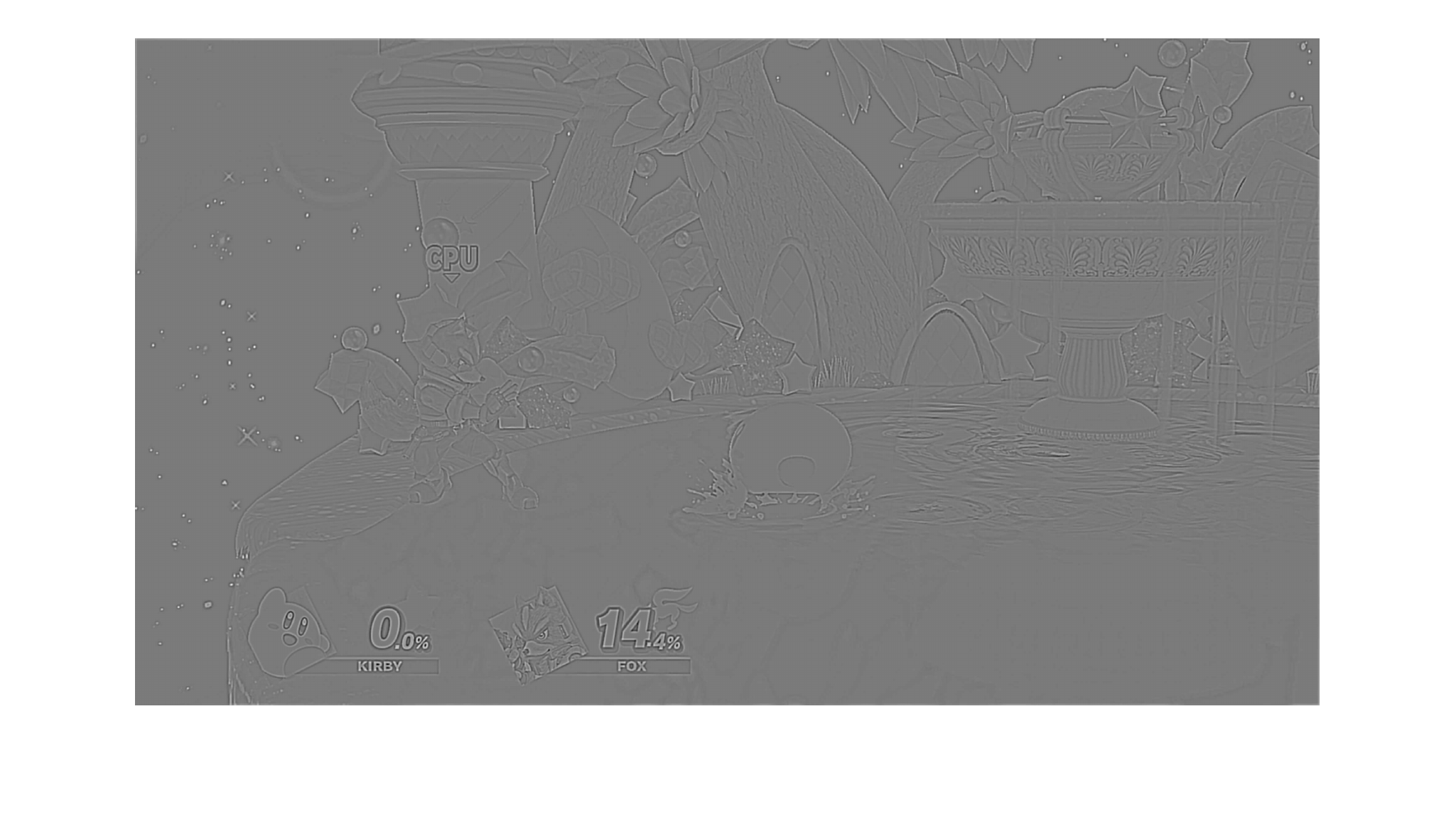}}
 \subfigure[]{
  \label{map_GM} 
  \includegraphics[width=0.23\textwidth]{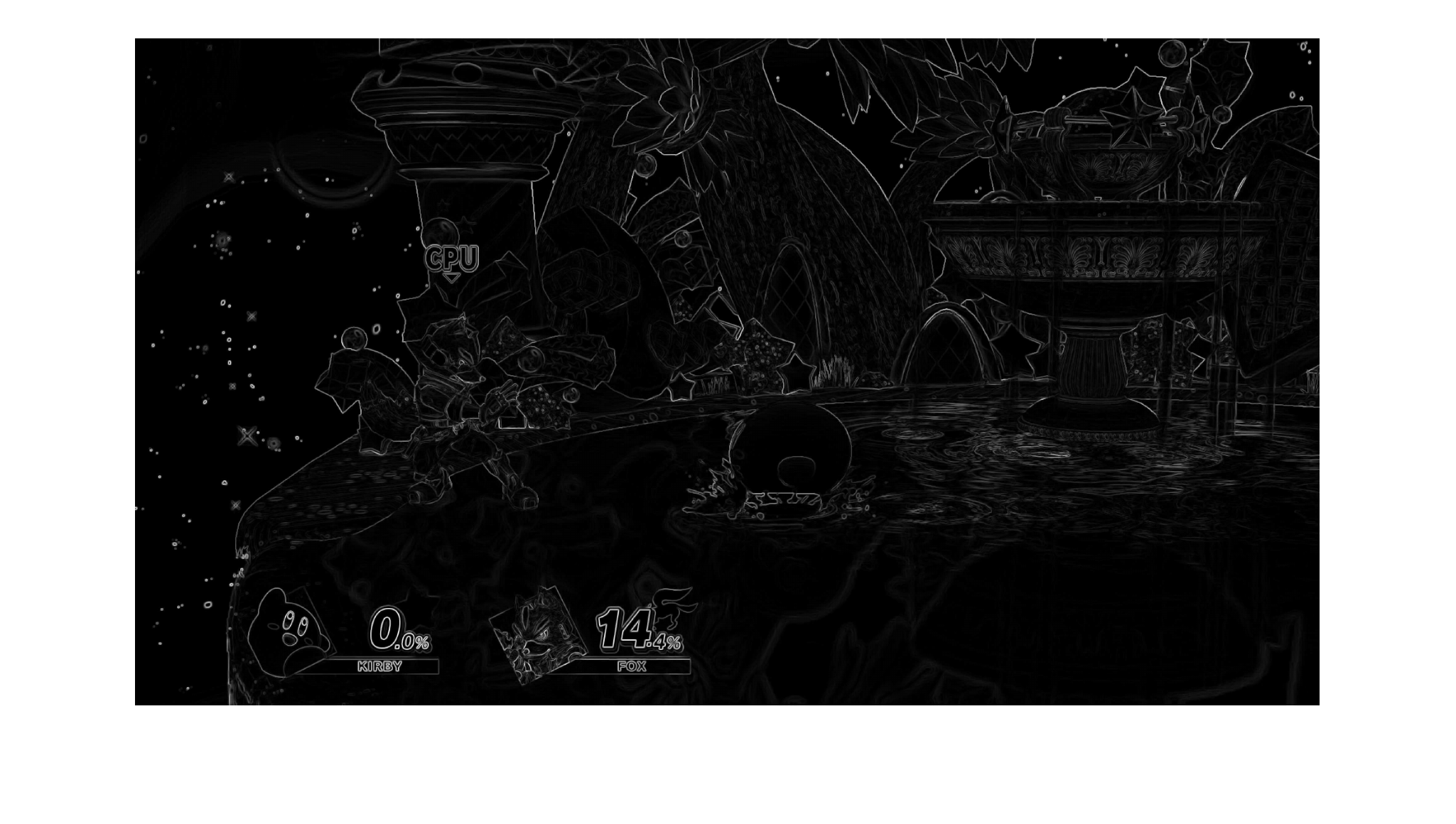}}   
 \subfigure[]{
  \label{map_diff} 
  \includegraphics[width=0.23\textwidth]{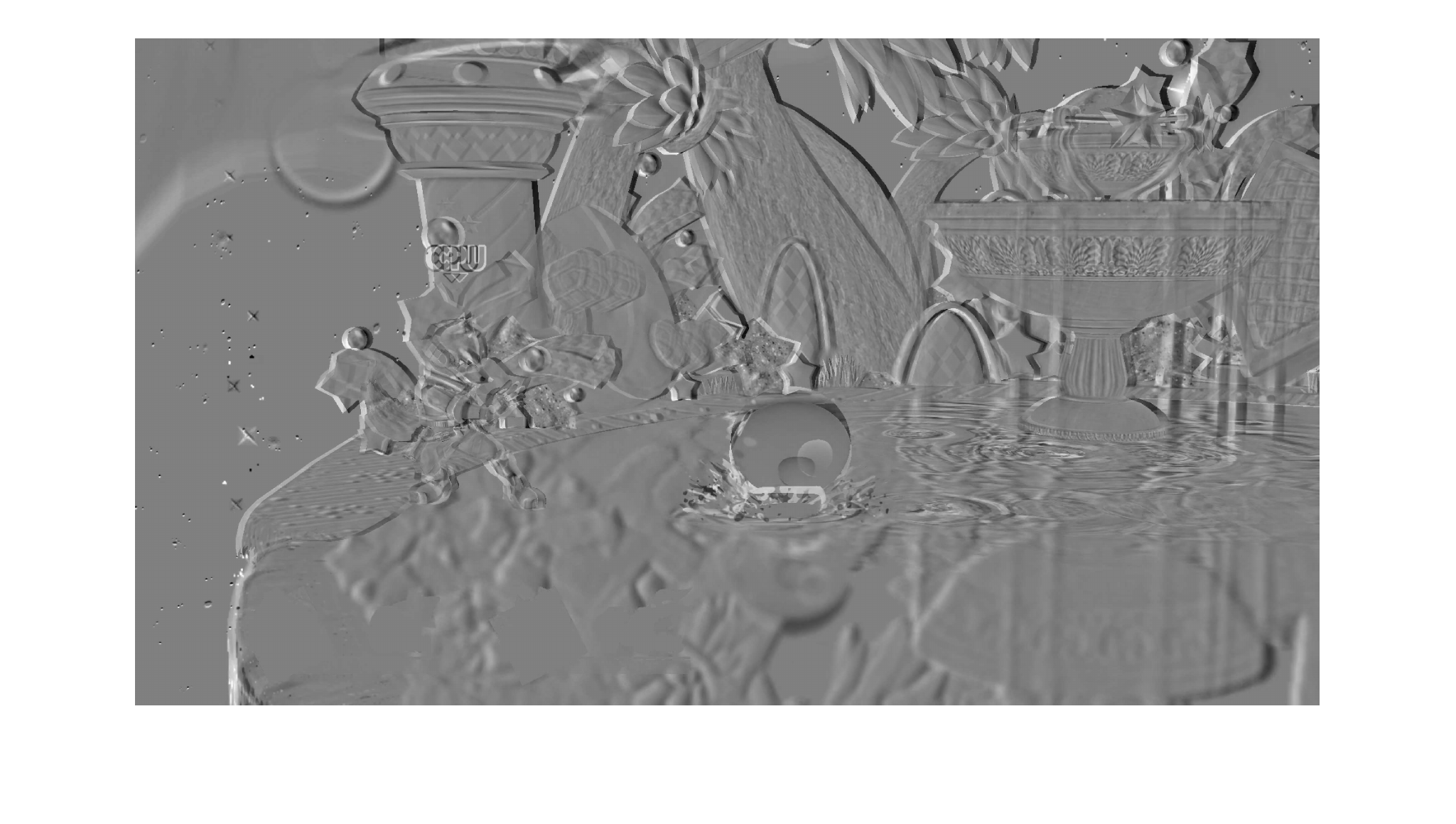}}         
 \caption{Color maps and processed coefficient maps of a gaming video taken from the game `Super Smash Bros' from the LIVE-YT-Gaming Database. (a) sRGB color map. (b) Luminance map $L^{*}$. (c) Chroma saturation map $C^{*}$. (d) Sigma field. (e) DoG field. (f) GM field. (g) Frame difference. }
\label{map_all} 
\end{figure*}

Many NSS-based algorithms are predicted on the assumption that the distributions of coefficients obtained by applying local (non-linear) bandpass filters followed by localized divisive energy normalization on natural images strongly tends towards a gaussian characteristic. 
When distortions occur, this tendency is disrupted, often in predictable ways. 
Although, as we shall show, synthetically generated gaming content does not follow the same statistical regularities as do natural images or videos, it is nevertheless regular. 
Likewise, these ``gaming video statistics'' (GVS) are also altered by the presence of distortions. 

An overview of the processing flow of the NSS module of GAME-VQP is shown in Fig. \ref{NSS_overview}. 
The flow is accomplished in five phases: color conversion, pre-processing, normalization, post-processing, and feature extraction. 

\subsubsection{Color Space/Map Phase}
The first phase is conversion to a suitable color space. 
There are many possibilities (RGB, YUV, LAB, etc.) and some VQA models use several color spaces \cite{ghadiyaram2017perceptual, tu2021rapique, chen2020perceptual, ghadiyaram2017capture, xue2014blind, zhang2015feature}. 
While most distortions manifest on the luminance channel, hence early NR VQA models discarded chromatic or color information \cite{mittal2012no, mittal2013making}, other distortions are more apparent in chromatic channels \cite{ghadiyaram2017perceptual, chen2020perceptual}. 
Since gaming video content is rich in color, and many distortions of gaming content affect color appearance, we deploy the perceptually uniform CIELCh space, which is derived from CIELAB, to compute luminance maps and chroma maps (\ref{chroma_map}) on which features are defined and computed in GAME-VQP. 
Specifically, two channels are used from CIELCh: the luminance channel $L$, which is identical to the luminance channel $L^{*}$ in CIELAB, representing perceived brightness, and the color saturation channel $C^{*}$, which is defined in terms of the chromatic channels $a^{*}$ and $b^{*}$ of CIELAB, which we convert to from original sRGB \cite{iec1999}. 
The LCh color space, which is given in polar form, has saturation values

\begin{equation}
\begin{aligned}
&C^{*}_{ab}=\sqrt{a^{*^{2}}+b^{*^{2}}}.\\
\label{chroma_map}
\end{aligned}
\end{equation}

\subsubsection{Pre-Processing Phase}
Fig. \ref{map_all} shows examples of the pre-processing methods as applied to an exemplar video frame. 
The `identity' blocks serve as placeholders by which some or all of the inputs are passed through without modification to the next phase of processing - as explained in the following. 
The second and third phases of processing prepare the videos for the extraction of perceptually-relevant, `quality-aware', statistical video features. 
The second phase of processing includes: identity, difference of Gaussian (DoG) filtering (\ref{DoG_equation}), gradient magnitude (GM) (\ref{eq:gm}) estimation, displaced frame differencing (DFD) (\ref{displaced_frame_diff}), and calculation of the `sigma field' (SF) (\ref{MSCN_var}) of each frame. 
We will define, explain the significance of, and detail how and where each of these pre-processing steps are applied, in the foregoing. 

The DoG filter is a broadly accepted model of the multi-scale receptive fields of retinal ganglion cells to visual stimuli. 
Moreover, it strongly enhances luminance/chroma edges, which are strongly present in synthetically generated gaming videos, and which are often degraded by common distortions. 
The spatial impulse response of a DoG filter takes the form: 
\begin{equation}
\begin{aligned}
&DoG(x, y)=\frac{1}{\sqrt{2\pi }}(\frac{1}{\sigma_{1}}e^{\frac{-(x^{2}+y^{2})}{2\sigma_{1}^{2}}}-\frac{1}{\sigma_{2}}e^{\frac{-(x^{2}+y^{2})}{2\sigma_{2}^{2}}})\\
\label{DoG_equation}
\end{aligned}
\end{equation}
\noindent
where $\sigma_{2}$ = $1.5\sigma_{1}$. 
The frequency response of (\ref{DoG_equation}) is a toroidal bandpass signature. 
The value of $\sigma_{1}$ used in our implementation was 1.16, following \cite{ghadiyaram2017perceptual}. 

We estimate the GM using a discrete sobel filter, using the orthogonal discrete convolution templates
\begin{equation}
\label{eq:sobel}
h_x=
\begin{bmatrix}
+1 & 0 & -1 \\
+2 & 0 & -2 \\
+1 & 0 & -1 \\
\end{bmatrix}
\ \text{and}\  
h_y=
\begin{bmatrix}
+1 & +2 & +1 \\
0 & 0 & 0 \\
-1 & -2 & -1 \\
\end{bmatrix},
\end{equation}
whereby the GM pixel map $I(i,j)$ is calculated as:
\begin{equation}
\label{eq:gm}
GM=\sqrt{(I\ast h_x)^2+(I\ast h_y)^2}.
\end{equation}
The GM responses are complementary to those of the DoG, as they provide less smooth, detailed information. 

Frame difference signals are known to possess highly regular statistical properties \cite{soundararajan2012video}, and contain strong indications of temporal distortions. 
This approach can be greatly enhanced by instead computing spatially displaced frame differences, the efficacy of which was demonstrated in the 1stepVQA \cite{yu2020predicting} model. 
Specifically, neighboring frames are spatially shifted in multiple directions before computing a corresponding set of DFD signals. 
Formally, given a video having $T$ luminance frames $I_{1}, I_{2}..., I_{t}..., I_{T}$, at each frame instant $t \in \{2,..., T\}$, compute 9 different DFD signals: 
\begin{equation}
\begin{aligned}
D_{tk}(i,j)=I_{t}(i,j)-I_{t+1}(i-k,j-l)
\label{displaced_frame_diff}
\end{aligned}
\end{equation}
\noindent
where $(k,l) \in \{(0, 0), (0, 1), (1, 0), (0, -1), (-1, 0), (-1, 1),\\ (1, -1), (-1, -1), (1, 1)  \}$ indexes the eight adjacent directions and one non-shift. 
The displaced frame differences provide simple but powerful descriptors of directional motion, which are highly regular along the direction of motion, and less so in other directions \cite{lee2020}. 

The sigma map is defined in Eq. (\ref{MSCN_var}) below, since it is also used in the next normalization phase. 
The sigma field captures regions of high activity, which are implicated in contrast masking processes \cite{bovik2013automatic} that are critical to distortion perception. 
Once computed, the sigma field is further enhanced by applying the DoG filter (\ref{DoG_equation}) on it. 
This bandpass process serves to substantially reduce the entropy of the sigma field, while highlighting the sparse set of high contrast elements, a strategy demonstrated in \cite{ghadiyaram2017perceptual, tu2021rapique} to significantly elevate quality prediction performance. 
Having defined five different types of pre-processing (including `identity' and 9 orientations of frame differencing) each is applied to both of the retained color maps ($L^{*}$ and $C^{*}$), yielding a total of 26 frame-sized processed color maps. 
Each of these is identically passed through the next ($3^{rd}$) processing phase of normalization. 

\subsubsection{Normalization Phase}
The third phase normalizes all ten of the pre-processed maps (including `identity') obtained in the first two phases by converting them into mean-subtracted, contrast-normalized (MSCN) coefficient maps (\ref{MSCN}). 
MSCN strongly de-correlates and normalizes the feature maps, desirable properties for learning and inferencing on. 
Originally devised to process luminance pictures in BRISQUE \cite{mittal2012no}, local mean subtraction is followed by divisive normalization on each map. 
Specifically, given each of the 26 maps produced in phase (2), indexed $P_{k}(i, j), k=1,...,26$, compute the MSCN coefficients: 
\begin{equation}
% \begin{aligned}
\hat{P}(i,j) = \frac{P(i,j) - \mu(i,j)}{\sigma(i,j) + C}
\label{MSCN}
% \end{aligned}
\end{equation}
where $i \in 1,2...M, j \in 1,2...N$ are spatial indices, and 
\begin{equation}
% \begin{aligned}
\mu(i,j) = \sum_{k = -K}^{K}\sum_{l = -L}^{L}w _{k,l}P(i-k,j-l),
\label{MSCN_mean}
% \end{aligned}
\end{equation}
\begin{equation}
% \begin{aligned}
\sigma(i,j) = \sqrt{\sum_{k = -K}^{K}\sum_{l = -L}^{L}w _{k,l}\big(P(i-k,j-l) - \mu(i,j)\big)^{2}}
\label{MSCN_var}
% \end{aligned}
\end{equation}
\noindent
are the weighted local mean and standard deviation of $P(i, j)$, where $C = 1$ is a stabilizing constant, and $w = \{w_{k,l}|k = -K,...,K, l = -L,...,L\}$ is a 2D circularly-symmetric Gaussian weighting function with $K = L = 3$. 

The normalization process (\ref{MSCN}) - (\ref{MSCN_var}) is a powerful regularizing operation on visual data. 
When applied on luminance or chrominance images that are not visibly distorted, it strongly decorrelates and gaussianizes the data \cite{mittal2012no, mittal2013making, ghadiyaram2017perceptual, ruderman1994statistics, xue2013gradient}. 
When applied on the four classes of pre-processed maps (other than `identity'), the data is similarly decorrelated and regularized. 
Distortions tend to predictably alter these characteristics by introducing correlation and model changes into the MSCN coefficients. 

\begin{figure*}
 \centering
 \footnotesize
 \subfigure[]{
  \label{filter_all:a}
  \includegraphics[width=0.23\textwidth]{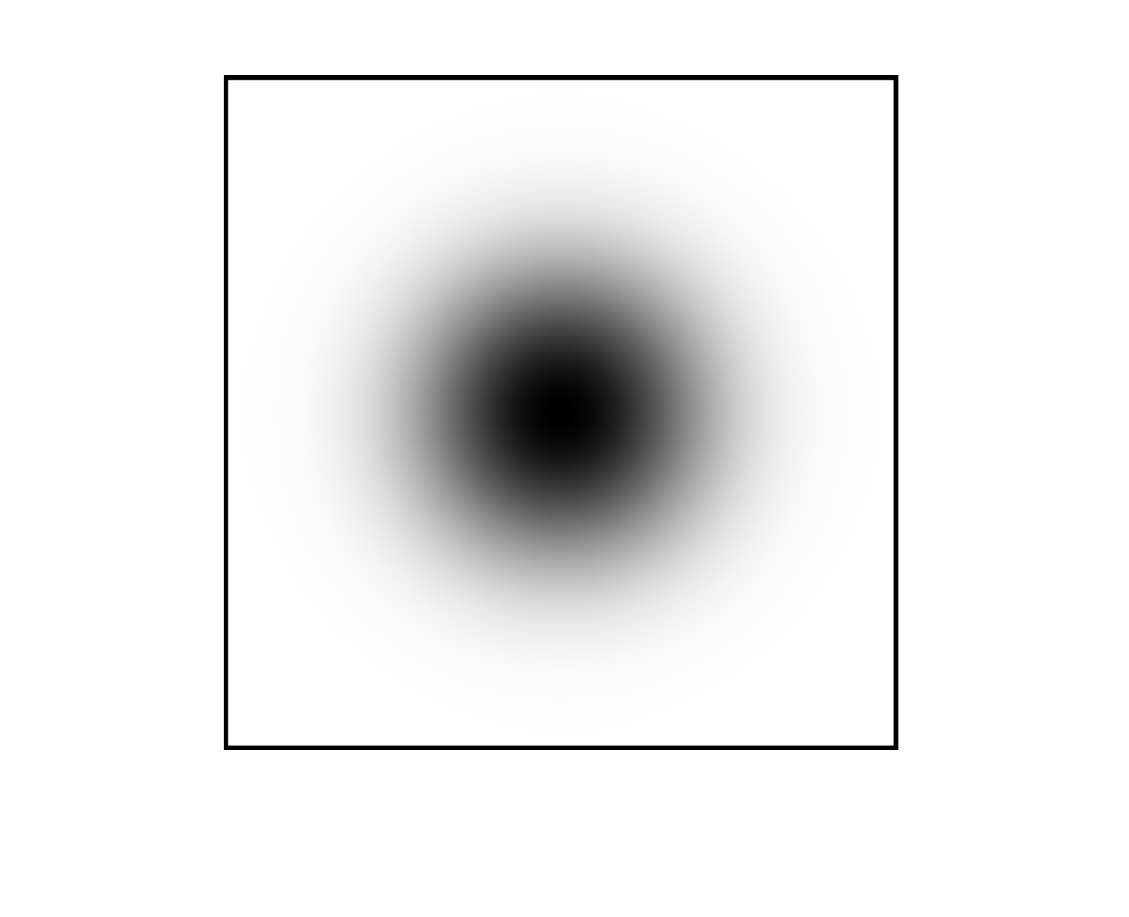}}
 \subfigure[]{
  \label{filter_all:b} 
  \includegraphics[width=0.23\textwidth]{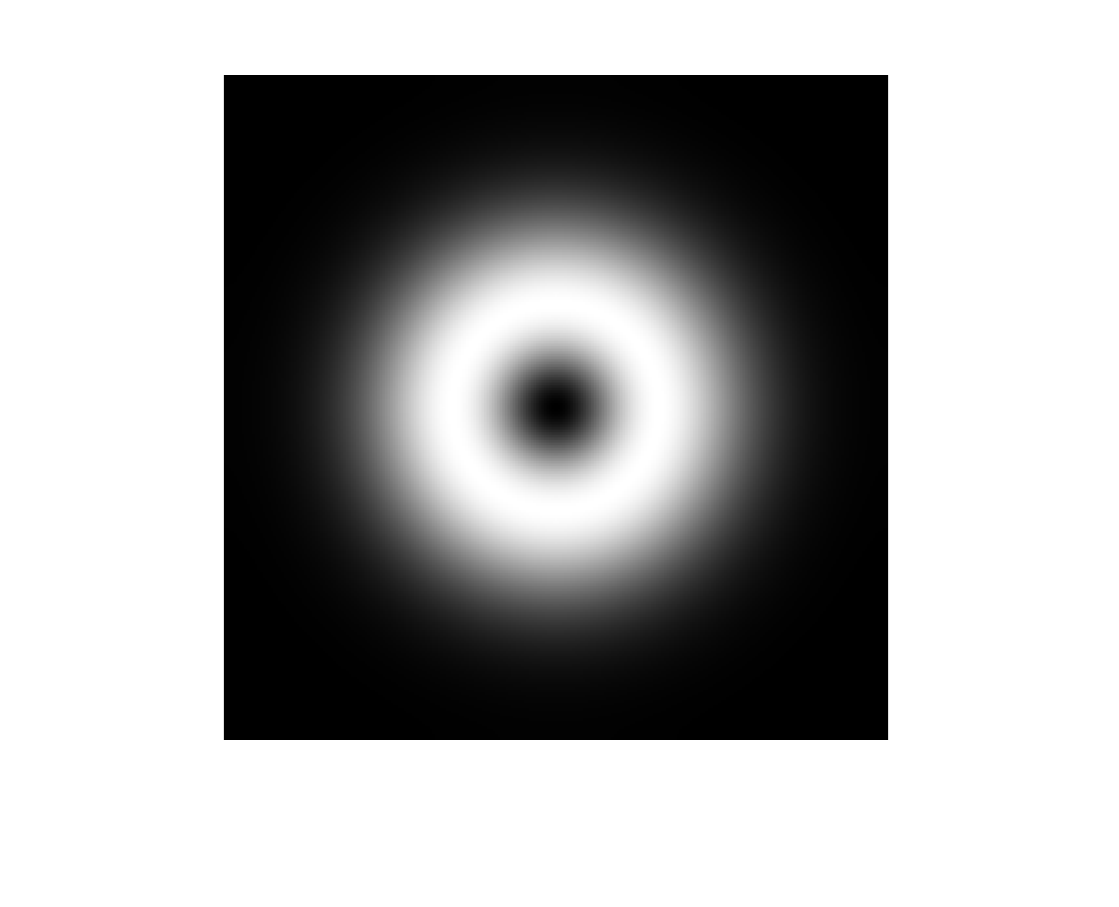}}   
 \subfigure[]{
  \label{filter_all:c} 
  \includegraphics[width=0.23\textwidth]{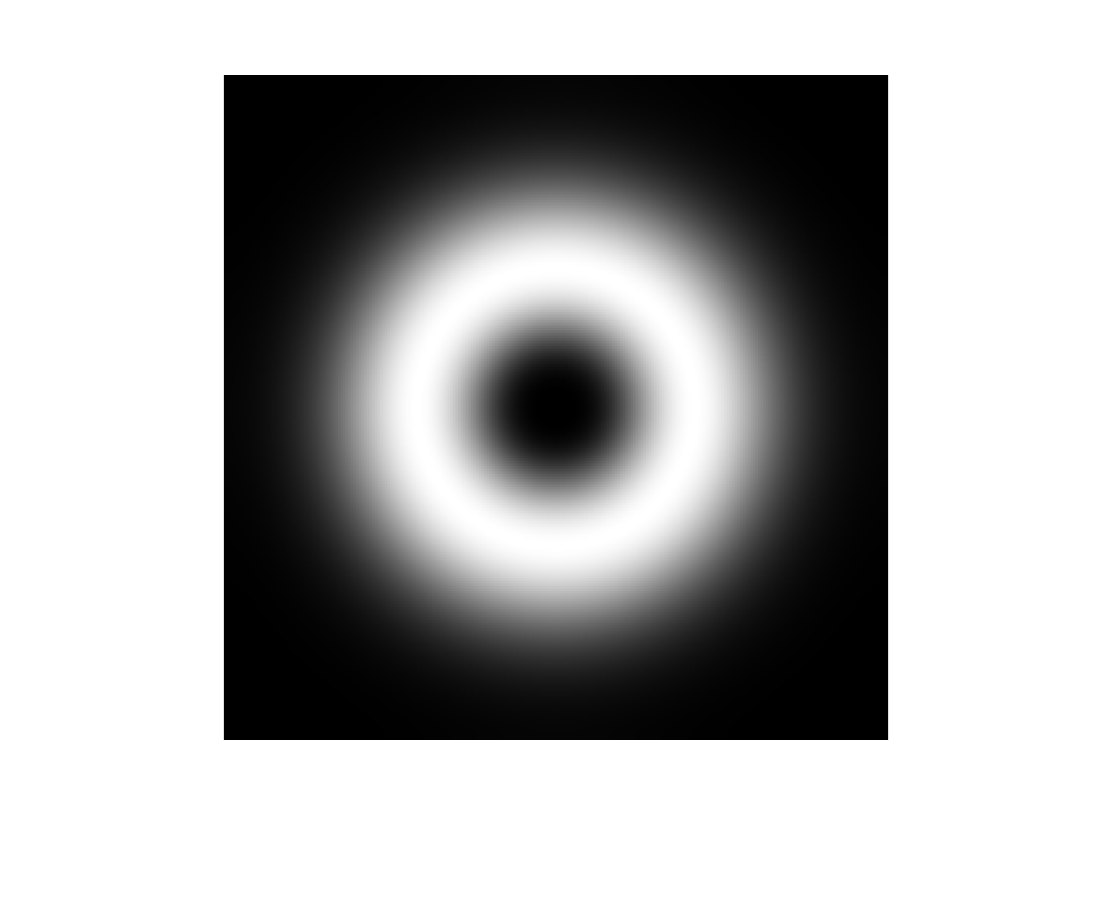}}      
 \caption{Discrete Fourier Transforms of (a) MS filter, (b) DoG filter (\ref{DoG_equation}); (c) convolution of (a) and (b). }
\label{filter_all} 
\end{figure*}

The question may arise that in two instances, consecutive DoG and MS (mean-subtraction) processes are applied. 
The aggregate of these two steps is a narrower bandpass operation than the DoG. 
As shown in Fig. \ref{filter_all:a}, MS is a high-pass operation, which when convolved with the bandpass DoG in Fig. \ref{filter_all:b}, produces a even narrower bandpass response (Fig. \ref{filter_all:c}). 
This yields stronger decorrelation than the DoG, and more regular statistical behavior, unless correlations and/or irregularities are introduced by distortion, which assists prediction. 

\begin{figure*}
 \centering
 \subfigure[]{
   \label{SuperSmashBros-012_200}
   \includegraphics[width=0.23\textwidth]{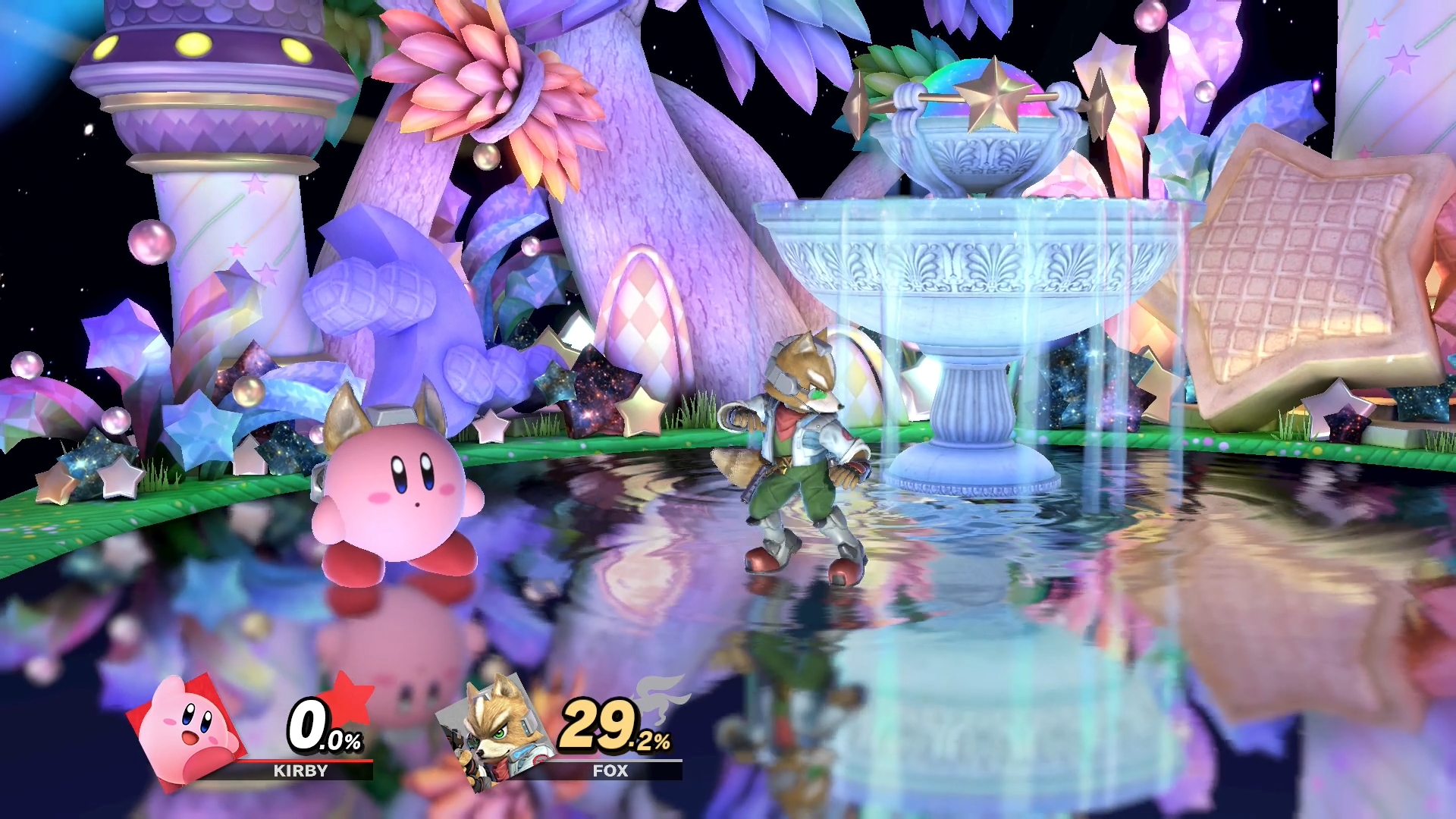}}
 \subfigure[]{
   \label{PUBG-089_200} 
   \includegraphics[width=0.23\textwidth]{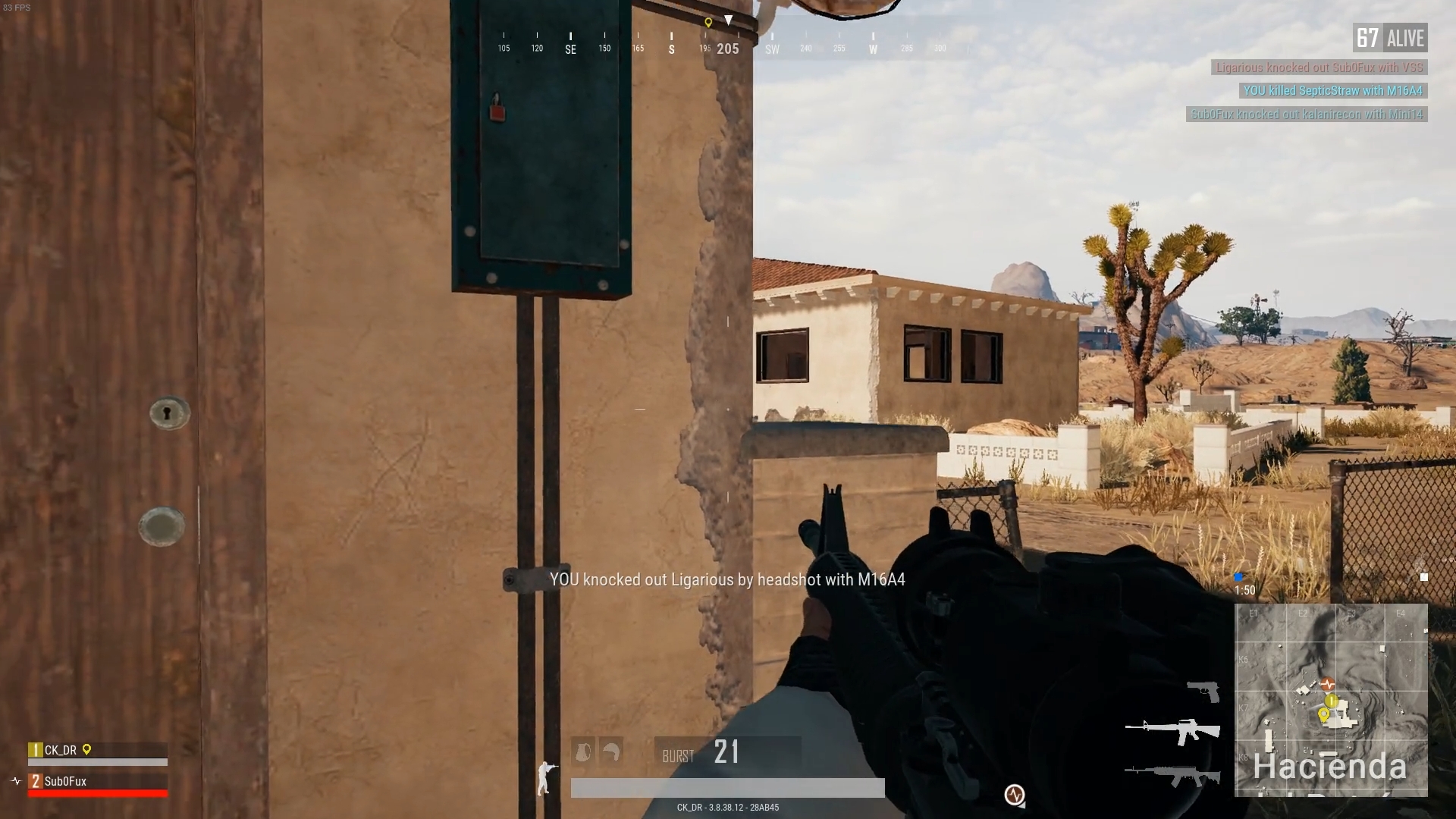}}
 \subfigure[]{
   \label{A073_150} 
   \includegraphics[width=0.23\textwidth]{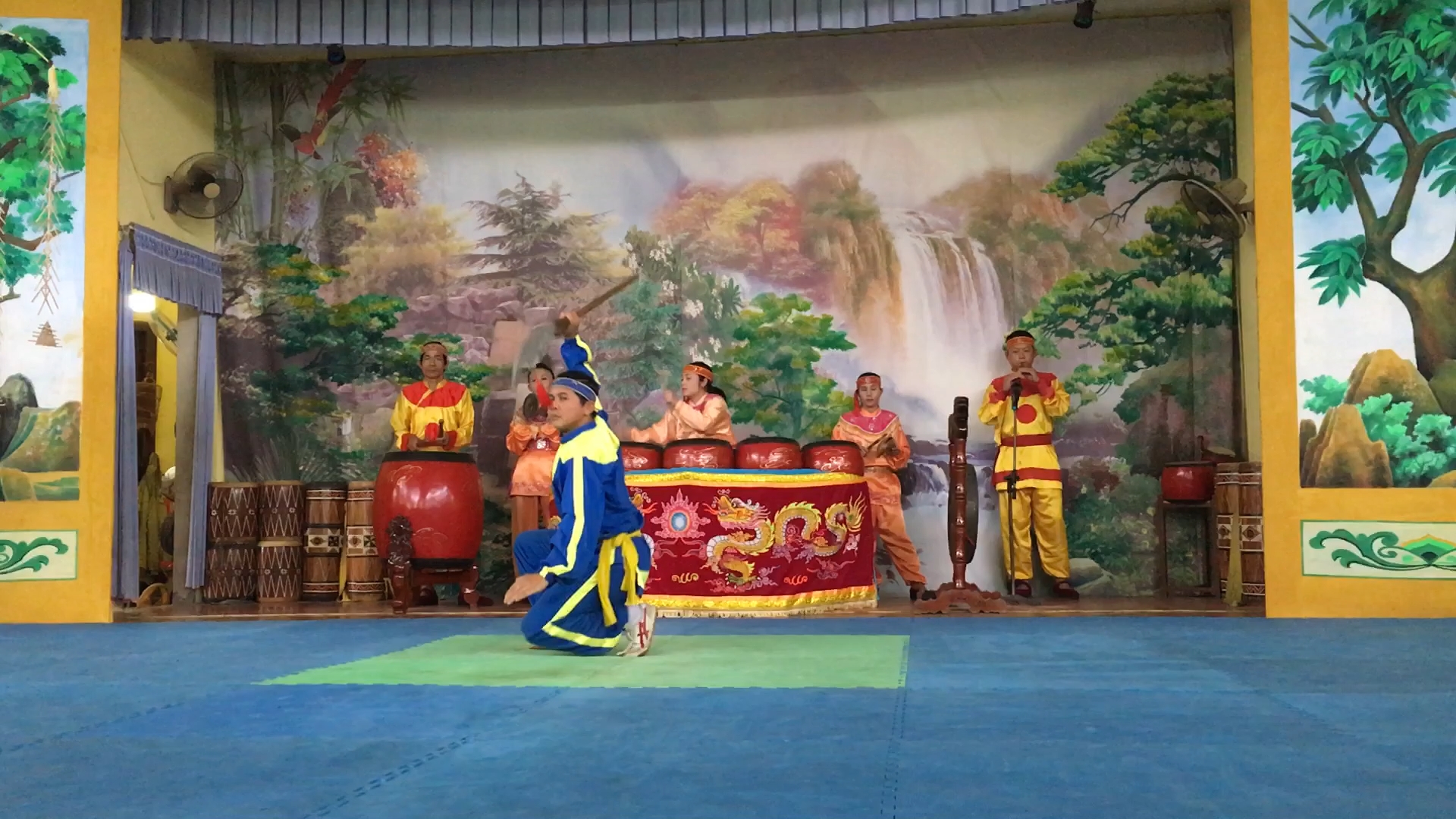}}   
 \subfigure[]{
   \label{B314_100} 
   \includegraphics[width=0.23\textwidth]{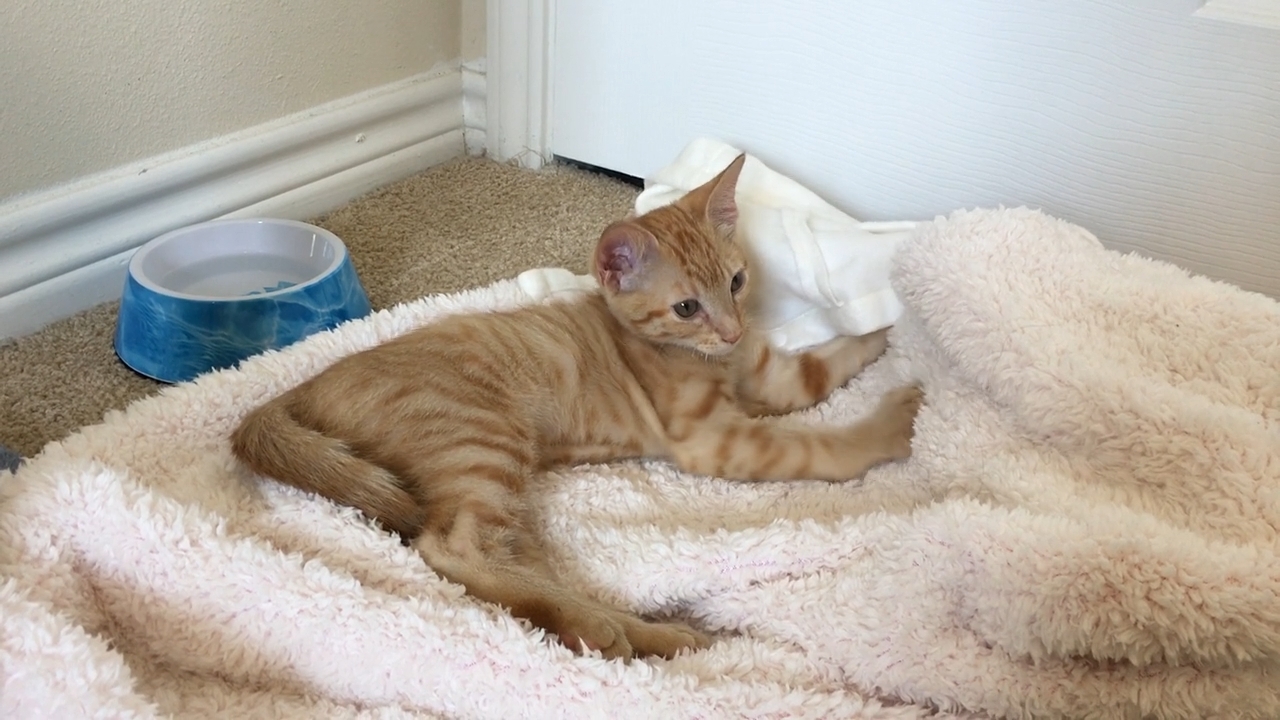}}      
 \caption{Screenshots of four exemplar videos. There are two gaming videos and two non-gaming videos that are compared. (a) Gaming video of `Super Smash Bros' from the LIVE-YT-Gaming Database (MOS: 95). (b) Gaming video of `PlayerUnknown's Battlegrounds (PUBG)' from the LIVE-YT-Gaming Database (MOS: 91). (c) Non-gaming video `A073' from the LIVE-VQC \cite{sinno2018large} Database (MOS: 91). (d) Non-gaming video `B314' from the LIVE-VQC Database (MOS: 94). }
\label{screenshot_all} 
\end{figure*}

\begin{figure}
\centering
\includegraphics[width = 1\columnwidth]{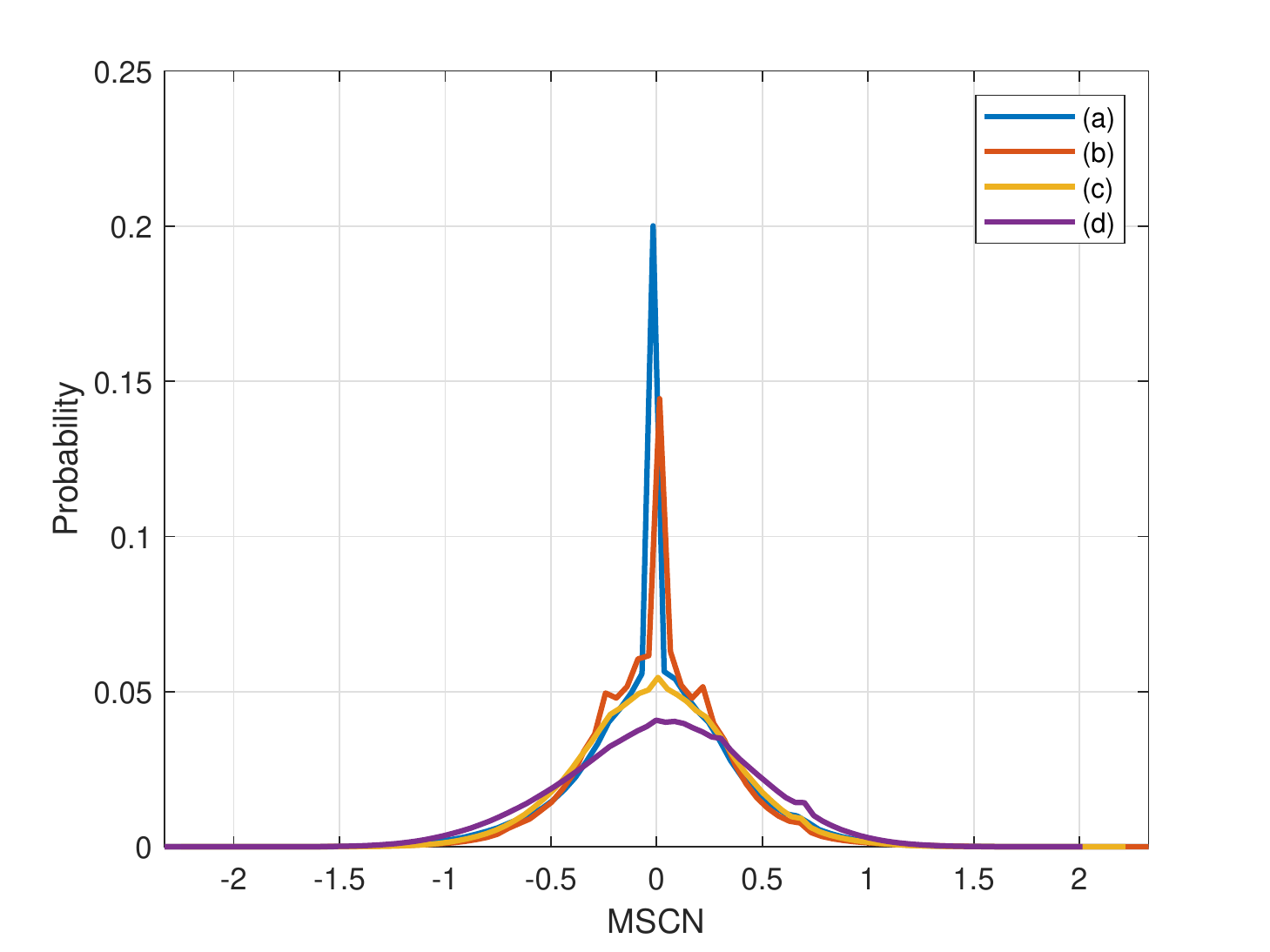}
\caption{MSCN coefficient histograms of the four videos (a)-(d) in Fig. \ref{screenshot_all}.}
\label{mscn_hist}
\end{figure}

We also studied whether there are meaningful statistical differences between the artificial gaming content and natural video content. 
To do this, we selected two high-quality gaming videos from the LIVE-YT-Gaming database and compared them with two high-quality, photographic UGC non-gaming videos selected from the LIVE-VQC \cite{sinno2018large} database. 
Screenshots of the four videos are shown in Fig. \ref{screenshot_all}, while Fig. \ref{mscn_hist} shows the MSCN coefficient distributions of the four videos. 
The MSCN coefficients of the two \textit{non}-gaming videos exhibit an obvious Gaussian distribution, while the distributions of the two gaming videos are much peakier and more heavy-tailed, similar to the Laplacian distribution. 
Although the four videos are all high-quality, the gaming videos do not conform to the traditional Gaussian statistical distribution underlying much of modern video quality theory. 
The likely reason for this is that synthetic game content tends to be much smoother in most places, except at edges which tend to be idealized. 
Moreover, synthetic illumination is also very smooth and tends not to alter these content properties, even if global illumination (GI) is used. 

\subsubsection{Post-Processing Phase}
The fourth phase allows for additional processing of the 26 MSCN maps.  
The purpose of this stage is to access the spatial correlation structure of some of the MSCN maps. 
It is not necessary to capture this kind of information on all of the 26 maps, and we found it to be adequate to compute the difference signals on the MSCN coefficients of the original `identity' images $L^{*}$ and $C^{*}$ maps, as on the GM's of them. 
Following computation of the MSCN coefficient maps, we analyze the statistics of the local spatial structure of the MSCN maps of $L^{*}$ and $C^{*}$ and the GM maps of $L^{*}$ and $C^{*}$. 
Given an MSCN map $P(i,j)$, four directional spatial differences are computed on it according to the following formulas: 
\begin{align}
\label{log_deri}
\begin{split}
& \nabla_1 P(i,j)  =P(i,j+1)-P(i,j) \\
& \nabla_2 P(i,j)  =P(i+1,j)-P(i,j)   \\
& \nabla_{3} P(i,j)  =P(i+1,j+1)-P(i,j) \\
& \nabla_{4} P(i,j)  =P(i+1,j-1)-P(i,j) \\
\end{split}
\end{align}

These quantities capture differential changes of the MSCN maps along several directions. 
Again, these are only applied on a subset of the MSCN maps. 
This results in 16 additional feature maps. 

\subsubsection{Feature Extraction Phase}
At this point we have computed 42 processed coefficient maps on which we will compute statistical features to inference quality predictions. 
The features are extracted under a parametric probability model that is fitted to the histograms of the 42 maps. 
The probability model we use is the classic Generalized Gaussian Distribution (GGD) which has been extensively used by to create highly effective picture/video quality predictors. 
It is a `sparsity distribution' that serves as a highly flexible model of bandpass, entropy-reduced video data. 

A zero mean GGD is given by: 
\begin{equation}
\begin{aligned}
f(x;\alpha ,\sigma ^{2})=\frac{\alpha }{2\beta \Gamma (1/\alpha)}exp(-(\frac{|x|}{\beta })^{\alpha })
\label{GGD}
\end{aligned}
\end{equation}
where
\begin{equation}
\begin{aligned}
\beta =\sigma \sqrt{\frac{\Gamma (1/\alpha)}{\Gamma(3/\alpha)}}
\label{GGD_beta}
\end{aligned}
\end{equation}
and $\Gamma(\cdot)$ is the gamma function:
\begin{equation}
\begin{aligned}
\Gamma (\alpha)=\int_{0}^{\infty}t^{\alpha -1}e^{-t}dt   \quad  a>0.
\label{GGD_gamma}
\end{aligned}
\end{equation}
Two parameters of the GGD are computed as features on all 42 processed coefficient maps: the shape parameter $\alpha$ which captures the degree of kurtotic behavior of the distribution, and the spread parameter $\sigma$. 
These are estimated using the popular and efficient moment-matching approach described in \cite{sharifi1995estimation}, resulting in a total of 84 features. 

\subsubsection{Summary of NSS Features}

\begin{table*}
\centering
\caption{Summary of Features Used in The GAME-VQP Model. }
\label{feature_list_table}
\resizebox{\textwidth}{!}{%
\begin{tabular}{|c|c|c|c|c|}
\hline
Color Space         & Pre-Processing              & Normalization & Post-Processing     & Number of $\alpha$, $\sigma$ Features \\ \hline
$L^{*}$ and $C^{*}$ & Identity  & MSCN          & -                   & 4                                     \\ \hline
$L^{*}$ and $C^{*}$ & DoG                         & MSCN          & -                   & 4                                     \\ \hline
$L^{*}$ and $C^{*}$ & Sigma-DoG                   & MSCN          & -                   & 4                                     \\ \hline
$L^{*}$ and $C^{*}$ & Displaced Frame Differences & MSCN          & -                   & 36                                    \\ \hline
$L^{*}$ and $C^{*}$ & Identity  & MSCN          & Spatial Differences & 16                                    \\ \hline
$L^{*}$ and $C^{*}$ & GM                          & MSCN          & -                   & 4                                     \\ \hline
$L^{*}$ and $C^{*}$ & GM                          & MSCN          & Spatial Differences & 16                                    \\ \hline
\end{tabular}
}
\end{table*}

Table \ref{feature_list_table} summarizes the color maps, pre-processing stages, normalization, post-processing, and GGD feature counts. 

\subsubsection{Multiscale Processing}

Finally, to capture the multi-scale attributes of videos and distortions of them, we repeat the entire process depicted in Fig. \ref{NSS_overview} and Table \ref{feature_list_table} on a spatially downscaled version of each video. 
The lower resolution video is obtained by spatially downscaling each frame by a factor of 2 along each dimension, using a suitable Gaussian anti-aliasing kernel identical to that used in BRISQUE \cite{mittal2012no}, NIQE, and other models. 
This doubles the feature count to 168. 

\subsection{CNN Module}

\begin{figure}
\centering
\includegraphics[width = 1\columnwidth]{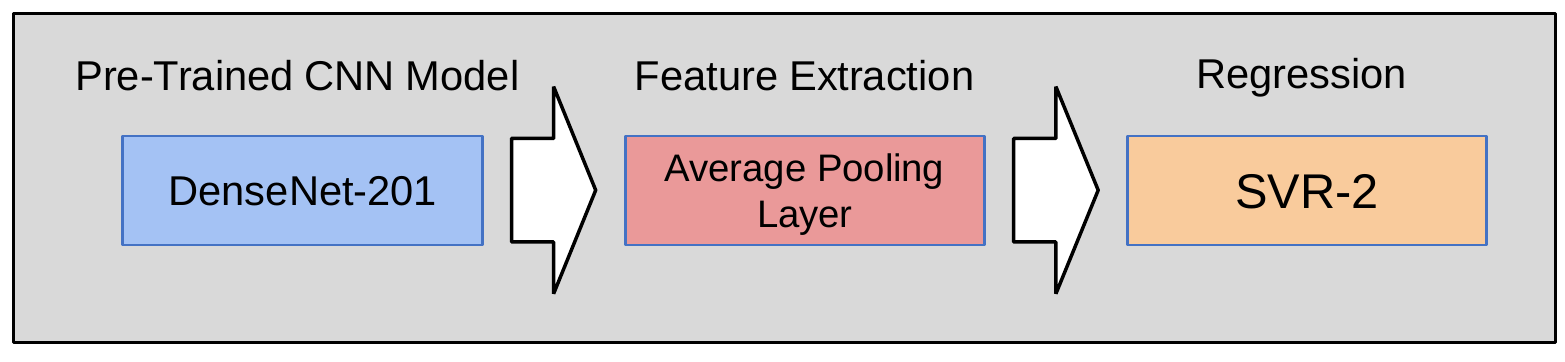}
\caption{Detailed flow chart of the CNN module.}
\label{CNN_overview}
\end{figure}

In addition to the statistical features, which only capture low-level attributes of video distortion, GAME-VQP is designed to also utilize features computed by a pre-trained CNN model on the raw RGB video frames. 
The purpose of the CNN model, which was pre-trained on the ImageNet \cite{deng2009imagenet} object recognition database, is provide complementary high-level, semantic features to the back-end inferencing engine. 
The CNN model in GAME-VQP uses a DenseNet CNN backbone feeding an average pooling layer, which supplies the feature outputs. 
Express the output feature vector thereby obtained as: 
\begin{equation}
\begin{aligned}
f_{l}=CNN(I^{'}(i,j)),
\label{CNN_FC}
\end{aligned}
\end{equation}
\noindent
where $CNN$ could be any CNN classification model, and $I^{'}$ represents the video being processed following spatial rescaling to fit the input size of the CNN. 
Since we are using the CNN to inference on high-level content attributes, resizing the inputs to the CNN models should not affect the pure quality aspects of prediction. 
In our implementation, we deployed a pre-trained DenseNet-201 CNN backbone on the video frames, and extracted the outputs of the final average pooling layer as the semantic features used for quality regression. 
No retraining of the CNN was done (the weights were frozen). 
The video was fed to the CNN in RGB format resized to 224x224x3 to fit the DenseNet. 
DenseNet-201 is a highly efficient residual network that has been shown to deliver high performance on the PGC gaming VQA problem \cite{utke2020ndnetgaming}. 
An overview of the CNN feature extraction network used in GAME-VQP is shown in Fig. \ref{CNN_overview}.

\subsection{Fusion Module}

Unlike many previous quality prediction models which directly feed all features as a single input vector into a regression engine, GAME-VQP processes the NSS and CNN features separately. 
The idea behind this approach is that, while content may affect quality perception (e.g., distorted faces are more annoying), the semantic meaning of the content is independent of distortion. 
Specifically, given a training or test video, compute all 168+1920 NSS and CNN features on each frame, then average each feature across all frames of the video to obtain two feature vectors of lengths 168 and 1920, respectively \cite{tu2020comparative}. 
The length-168 NSS feature vector is used to train and inference with one model (denoted SVR-1); while the length-1920 CNN feature vector supplies inputs to the second, independent SVR model (SVR-2). 
Thus, SVR-1 produces quality predictions based on true `quality-aware' input features, while SVR-2 produces predictions of the degree to which learned semantics relate to perceived video quality. 
Since these aspects are likely cumulative to the overall perception of quality (i.e., of distortion and of content relevancy to quality), in the end the arithmetic mean value of the responses obtained from the two SVRs yields the final predicted video quality scores: 
\begin{equation}
\begin{aligned}
Q_{GAME-VQP}=\frac{M_{1}+M_{2}}{2},
\label{fusion_equation}
\end{aligned}
\end{equation}
where $M_{i}$ is the output of SVR-$i, i=1, 2$. 

\section{Performance Evaluation}
\label{performance_evaluation}

We conducted extensive experiments on the LIVE-YT-Gaming database (Section \ref{LIVE-YT-Gaming_intro}) to study the performance of GAME-VQP against that of existing VQA algorithms. 
We selected several popular and powerful mainstream, general-purpose NR VQA algorithms for comparison, along with the current best performing NR VQA algorithm specifically designed for gaming VQA. 
We selected four general-purpose NR VQA algorithms: BRISQUE, TLVQM, VIDEVAL, and RAPIQUE (based on hand-selected features), a deep learning-based algorithm called VSFA, a simple but powerful learning-free `completely blind' model called NIQE, and a very recent deep learning algorithm called NDNetGaming designed for gaming videos. 
GAME-VQP, BRISQUE, TLVQM, VIDEVAL, and RAPIQUE all extract video features first, then perform regression. 
For fair comparison, we utilized the same model and configuration of SVR as is widely used in many previous designs \cite{mittal2012no, ghadiyaram2017perceptual, tu2021rapique}. 
When testing TLVQM, VIDEVAL, and RAPIQUE which are true space-time VQA models, we used the default temporal pooling method of each algorithm to arrive at a single feature vector for each video. 
We used the Matlab LIBSVM package to implement the SVR models. 
For the deep models VSFA and NDNetGaming, we implemented the original releases published by the authors. 

We used three criteria to evaluate the performances of the compared algorithms: the Spearman’s rank order correlation coefficient (SROCC), the Pearson’s (linear) correlation coefficient (LCC), and the Root Mean Squared Error (RMSE). 
In each case the correlations are taken between model predictions and the corresponding MOS in the LIVE-YT-Gaming Video Quality Database. 
Before calculating LCC and RMSE, the predicted quality scores were fitted to a 4 parameter logistic function, as recommended in \cite{VQEG2000}. 
Larger values of SROCC and LCC imply better performance, while larger values of RMSE indicate worse performance. 

To make fair comparisons between the VQA algorithms, we randomly divided the database into a non-overlapping 80\% training set containing 480 videos and a 20\% test set containing 120 videos. 
To avoid biased results, we repeated this random splitting process over 100 iterations, and report the median value of the performance results. 
NIQE does not require training, so we simply report its results on the test subset. 

\subsection{Algorithm Comparisons}

\begin{table}
\caption{Results of One-Sided Wilcoxon Rank Sum Test Performed Between SROCC Values of The Compared VQA Algorithms In Table \ref{performance_model}. A Value Of "1" Indicates That The Row Algorithm Was Statistically Superior to The Column Algorithm; " $-$ 1" Indicates That the Row Was Worse Than the Column; A Value Of "0" Indicates That the Two Algorithms Were Statistically Indistinguishable. }
\label{performance_statistc_srocc}
\resizebox{\columnwidth}{!}{%
\begin{tabular}{|c|c|c|c|c|c|c|c|c|}
\hline
            & NIQE & BRISQUE & TLVQM & VIDEVAL & RAPIQUE & VSFA & NDNetGaming & GAME-VQP \\ \hline
NIQE        & 0    & -1      & -1    & -1      & -1      & -1   & -1          & -1       \\ \hline
BRISQUE     & 1    & 0       & -1    & -1      & -1      & -1   & 1           & -1       \\ \hline
TLVQM       & 1    & 1       & 0     & -1      & -1      & -1   & 1           & -1       \\ \hline
VIDEVAL     & 1    & 1       & 1     & 0       & 0       & 1    & 1           & -1       \\ \hline
RAPIQUE     & 1    & 1       & 1     & 0       & 0       & 1    & 1           & -1       \\ \hline
VSFA        & 1    & 1       & 1     & -1      & -1      & 0    & 1           & -1       \\ \hline
NDNetGaming & 1    & -1      & -1    & -1      & -1      & -1   & 0           & -1       \\ \hline
GAME-VQP    & 1    & 1       & 1     & 1       & 1       & 1    & 1           & 0        \\ \hline
\end{tabular}
}
\end{table}

\begin{table}
\caption{Results of One-Sided Wilcoxon Rank Sum Test Performed Between LCC Values of The Compared VQA Algorithms In Table \ref{performance_model}. A Value Of "1" Indicates That The Row Algorithm Was Statistically Superior to The Column Algorithm; " $-$ 1" Indicates That the Row Was Worse Than the Column; A Value Of "0" Indicates That the Two Algorithms Were Statistically Indistinguishable. }
\label{performance_statistc_lcc}
\resizebox{\columnwidth}{!}{%
\begin{tabular}{|c|c|c|c|c|c|c|c|c|}
\hline
            & NIQE & BRISQUE & TLVQM & VIDEVAL & RAPIQUE & VSFA & NDNetGaming & GAME-VQP \\ \hline
NIQE        & 0    & -1      & -1    & -1      & -1      & -1   & -1          & -1       \\ \hline
BRISQUE     & 1    & 0       & -1    & -1      & -1      & -1   & 1           & -1       \\ \hline
TLVQM       & 1    & 1       & 0     & -1      & -1      & -1   & 1           & -1       \\ \hline
VIDEVAL     & 1    & 1       & 1     & 0       & -1      & 1    & 1           & -1       \\ \hline
RAPIQUE     & 1    & 1       & 1     & 1       & 0       & 1    & 1           & -1       \\ \hline
VSFA        & 1    & 1       & 1     & -1      & -1      & 0    & 1           & -1       \\ \hline
NDNetGaming & 1    & -1      & -1    & -1      & -1      & -1   & 0           & -1       \\ \hline
GAME-VQP    & 1    & 1       & 1     & 1       & 1       & 1    & 1           & 0        \\ \hline
\end{tabular}
}
\end{table}

\begin{table*}
\caption{Performances of Compared No-Reference VQA Models on The LIVE-YouTube Gaming Video Quality Database Using Randomly Chosen, Non-Overlapping 80\% Training and 20\% Test Sets. The Numbers Denote Median Values Over $100$ Such Randomly Divisions. Boldfaces Indicate The Top Performing Model in Each Row. Italics Indicate Deep Learned VQA Models. Underlines Indicate VQA Models Designed for Gaming Videos. }
\label{performance_model}
\resizebox{\textwidth}{!}{%
\begin{tabular}{|c|c|c|c|c|c|c|c|c|}
\hline
      & NIQE   & BRISQUE & TLVQM  & VIDEVAL & RAPIQUE & \textit{VSFA} & {\ul NDNetGaming} & {\ul \textbf{GAME-VQP}} \\ \hline
SROCC & 0.2801 & 0.6037  & 0.7484 & 0.8071  & 0.8028  & 0.7762        & 0.4562            & \textbf{0.8563}   \\ \hline
LCC   & 0.3037 & 0.6383  & 0.7564 & 0.8118  & 0.8248  & 0.8014        & 0.4690            & \textbf{0.8754}   \\ \hline
RMSE  & 16.208 & 16.208  & 11.134 & 10.093  & 9.661   & 10.396        & 14.941            & \textbf{8.533}    \\ \hline
\end{tabular}
}
\end{table*}

\begin{figure}
 \centering
 \subfigure[]{
   \label{performance_boxplot:SROCC}
   \includegraphics[width=0.8\columnwidth]{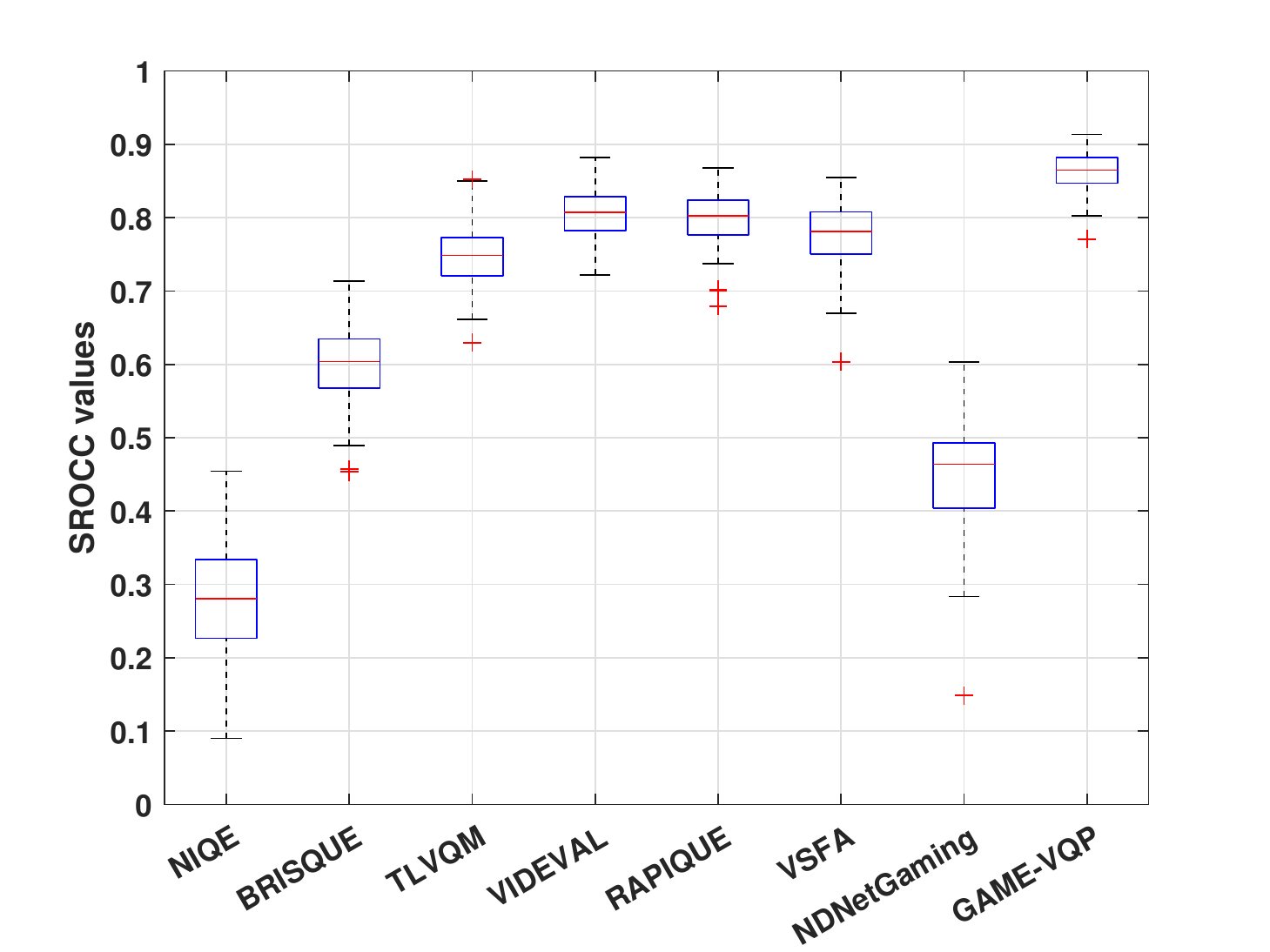}}
 \subfigure[]{
   \label{performance_boxplot:LCC} 
   \includegraphics[width=0.8\columnwidth]{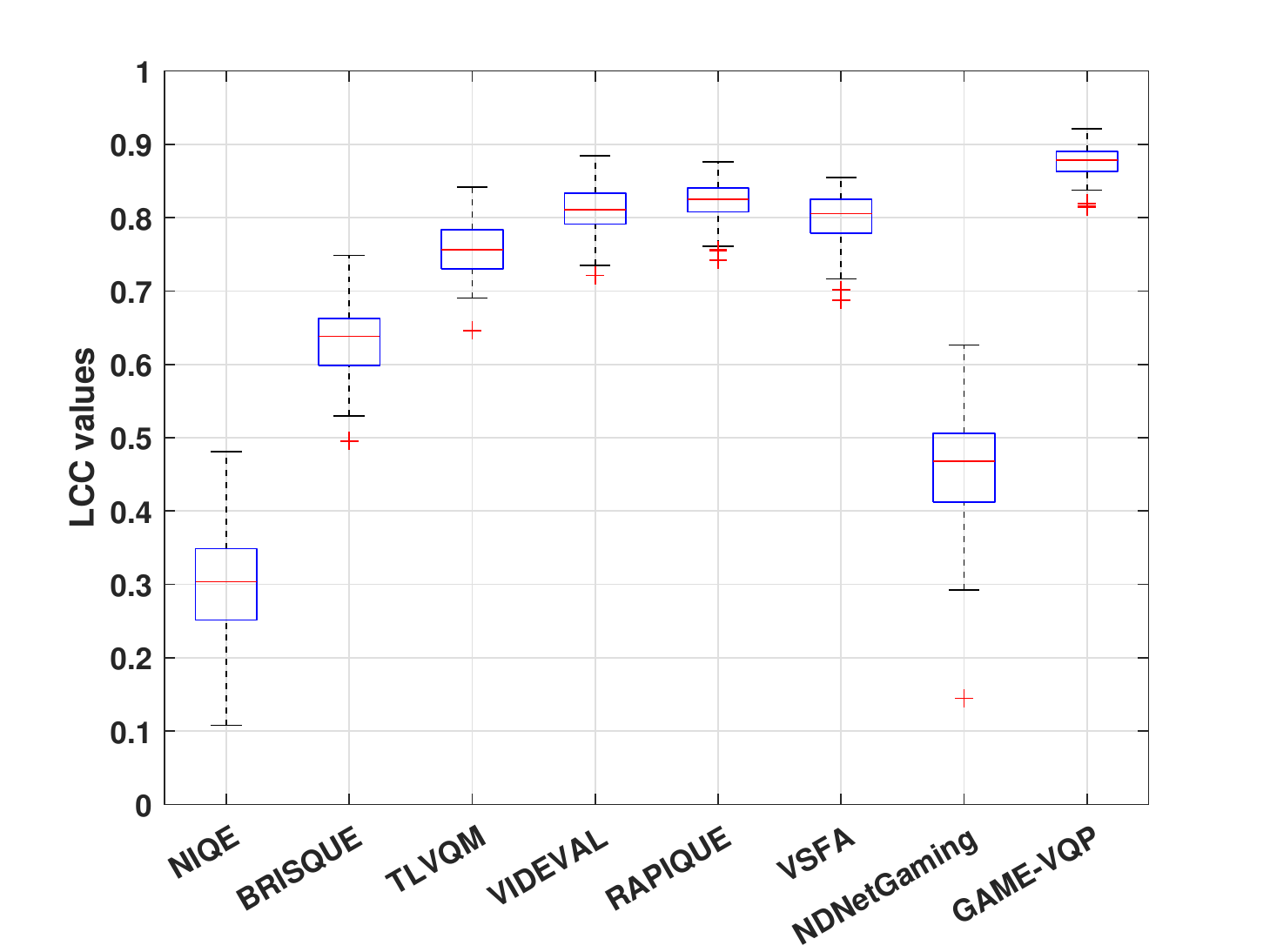}}
 \caption{Box plots of the SROCC and LCC distributions of the compared algorithms in Table \ref{performance_model}, over 100 randomized trails on the LIVE-YouTube Gaming Video Quality Database. The central red marks represent the median, while the bottom and top edges of the box indicate the 25th and 75th percentiles, respectively. The whiskers extend to the most extreme data points not considered outliers, while the outliers are plotted with red '+' symbols. }
 \label{performance_boxplot} 
\end{figure}

To determine whether there were significant differences in the performances of the compared models, we first performed a statistical significance test. 
We used the non-parametric Wilcoxon Rank Sum Test \cite{wilcoxon1945individual} to determine whether the distributions of SROCC and LCC values of the compared models, computed over 100 random train-test splits, are statistically equivalent or not. 
Tables \ref{performance_statistc_srocc} and \ref{performance_statistc_lcc} shows the results of the significance test, indicating that GAME-VQP performed significantly better than all of the other compared VQA algorithms in terms of SROCC and LCC, at the 95\% confidence level. 

Table \ref{performance_model} lists the SROCC, LCC, and RMSE performances of all of the compared models on the LIVE-YT-Gaming dataset. 
Clearly, the worst performer was NIQE, which is not surprising, given that it is a training-free, completely blind model. 
BRISQUE, which uses the same features as NIQE, yielded significantly better performance when retrained on the gaming video data and subjective scores. 
While both BRISQUE and NIQE have been shown to perform well on PGC gaming content impaired by single compression distortions \cite{barman2018objective}, they performed much worse on UGC gaming content. 
TLVQM, which computes motion-related statistical features, performed better than either of the frame-based models. 
VIDEVAL and RAPIQUE, which both use NSS features along with other features, produced further improved performances. 
RAPIQUE \cite{tu2021rapique} uses many features ($>$3000) along with deep features in an extremely efficient way, while VIDEVAL uses a greatly reduced feature set, but at much greater expense, to achieve similar results. 
The deep learning model VSFA deploys a Resnet-50 backbone, which feeds a GRU \cite{cho2014learning} to capture temporal quality characteristics of the videos. 
GAME-VQP was undoubtedly the best performing algorithm. 
Furthermore, GAME-VQP also produced the most stable prediction results, as may be observed from the boxplot in Fig. \ref{performance_boxplot}. 

\subsection{Comparison and Analysis of Pre-Trained CNN Models}
\label{CNN_analysis}

As an additional benchmark, we also tested four CNN networks originally used for image classification problems after being trained on more than a million images from the ImageNet database: VGG-16, Xception, Resnet-50, and DenseNet-201. 
VGG-16 and Resnet-50 have often been deployed in IQA and VQA models \cite{kim2017deep, tu2021rapique, tu2021ugc, li2019quality}. 
Xception and DenseNet-201 delivered better performance than the prior two deep models on the ImageNet classification problem. 
We applied these four pre-trained networks on suitably downscaled video frames, extracted feature vectors from their final average pooling or fully connected layers, and used them to train SVR models for each network. 
As may be seen from Table \ref{performance_CNN_networks}, when using the same train-test protocol and iteration count as before, the VGG-16 model performed the worst when tested on the LIVE-YT-Gaming dataset. 
The Xception and Resnet-50 models performed similarly, while the performance of the DenseNet-201 significantly surpassed that of the other networks, and was comparable to that of RAPIQUE and VIDEVAL. 
However, none of these models was able to approach the performance of GAME-VQP.

\begin{table}
\caption{Performance of Four Pre-Trained CNN Networks on the LIVE-YT-Gaming Database. Bold Font Represents the Best-Performing Network in Each of Row. }
\centering
\label{performance_CNN_networks}
\begin{tabular}{|c|c|c|c|c|}
\hline
      & VGG-16 & Xception & Resnet-50 & \textbf{DenseNet-201} \\ \hline
SROCC & 0.5768 & 0.7315   & 0.7290    & \textbf{0.8075}      \\ \hline
LCC   & 0.6429 & 0.7594   & 0.7677    & \textbf{0.8299}      \\ \hline
RMSE  & 13.240 & 11.160   & 11.083    & \textbf{9.692}       \\ \hline
\end{tabular}
\end{table}

\subsection{Comparison of Fusion Methods}

\begin{table}
\caption{Performance Comparison between Different Fusion Methods. Bold Font Represents the Best-Performing Network in Each of Row. }
\centering
\label{performance_combination}
\resizebox{\columnwidth}{!}{%
\begin{tabular}{|c|c|c|c|c|}
\hline
      & NSS    & Single SVR & \textbf{Mean Fusion} & \textbf{Product Fusion} \\ \hline
SROCC & 0.8094 & 0.8236     & \textbf{0.8563}         & \textbf{0.8546}         \\ \hline
LCC   & 0.8230 & 0.8469     & \textbf{0.8754}         & \textbf{0.8712}         \\ \hline
RMSE  & 9.794  & 9.354      & \textbf{8.533}          & \textbf{8.593}          \\ \hline
\end{tabular}
}
\end{table}

We compared the SVR prediction fusion method used in GAME-VQP (the mean of the SVR-1, and SVR-2 outputs) with two other methods. 
\begin{itemize}
  \item Single SVR: Stitch the extracted NSS and CNN features into a feature vector, and use them to train a single SVR model, instead of two. 
  \item Mean Fusion: Compute the mean of the predictions produced by SVR-1 and SVR-2, as currently employed in GAME-VQP. 
  \item Product Fusion: Compute the product of the predictions produced by SVR-1 and SVR-2 instead of the mean value. 
  \end{itemize}

As Table \ref{performance_combination} shows, using the product or mean of the SVR responses yielded essentially identical performances. 
Both, however, were much better than training the SVR to learn both NSS-quality and deep semantic features. 
Finally, we also trained a single SVR to map just the NSS features used by GAME-VQP. 
As it turns out, this NSS-only model performed similarly to VIDEVAL and RAPIQUE. 
By selecting features relevant to gaming videos, a model trained on NSS features is able to achieve performance comparable to the much more complex RAPIQUE.

\subsection{Performance on Non-Gaming UGC Videos}
\begin{table*}
\centering
  \begin{threeparttable}
\caption{Performance Comparison of Various No-Reference VQA Models on the Live Video Quality Challenge Database (LIVE-VQC). The Underlined And Boldfaced Entries Indicate the Best And Top Three Performers for Each Performance Metric in Each Row, Respectively. Except for the GAME-VQP Scores, the Results Are Cited From Experiments Reported in \cite{tu2021rapique}. }
\label{UGC_performance}
\begin{tabular}{|c|c|c|c|c|c|c|c|c|}
\hline
      & BRISQUE & {\ul \textbf{TLVQM}}  & VIDEVAL & \textbf{RAPIQUE} & VSFA   & GAME-VQP-NSS & GAME-VQP-CNN & \textbf{GAME-VQP} \\ \hline
SROCC & 0.5925  & {\ul \textbf{0.7988}} & 0.7522  & \textbf{0.7548}  & 0.6978 & 0.6987       & 0.6714       & \textbf{0.7549}   \\ \hline
LCC   & 0.6380  & {\ul \textbf{0.8025}} & 0.7514  & \textbf{0.7863}  & 0.7426 & 0.7176       & 0.7413       & \textbf{0.7933}   \\ \hline
RMSE  & 13.100  & {\ul \textbf{10.145}} & 11.100  & \textbf{10.518}  & 11.649 & 12.116       & 11.419       & \textbf{10.582}   \\ \hline
\end{tabular}
\end{threeparttable}
\end{table*}

While GAME-VQP is specifically designed to conduct VQA on gaming videos, it is of interest to study its performance on ordinary, non-gaming videos, to ascertain whether it is indeed ``specialized'' or is generalizable. 
Hence, we also conducted experiments on a large database of generic UGC videos. 
The LIVE Video Quality Challenge Database (LIVE-VQC) \cite{sinno2018large} contains unique UGC videos, and is widely used to validate the performances of NR VQA models. 
We compared the performance of GAME-VQP on LIVE-VQC against other SOTA algorithms with results shown in Table \ref{UGC_performance}. 
As may be observed, the predicted results of GAME-VQP were nearly as good as those of the two top-performing SOTA models on LIVE-VQC (TLVQM and RAPIQUE). 
This shows that GAME-VQP is indeed specialized, but also reasonably generalizeable to the broader UGC VQA problem.

\subsection{Scatter Plots}

\begin{figure}
 \centering
 \subfigure[]{
   \label{performance_scatter:NIQE}
   \includegraphics[width=0.48\columnwidth]{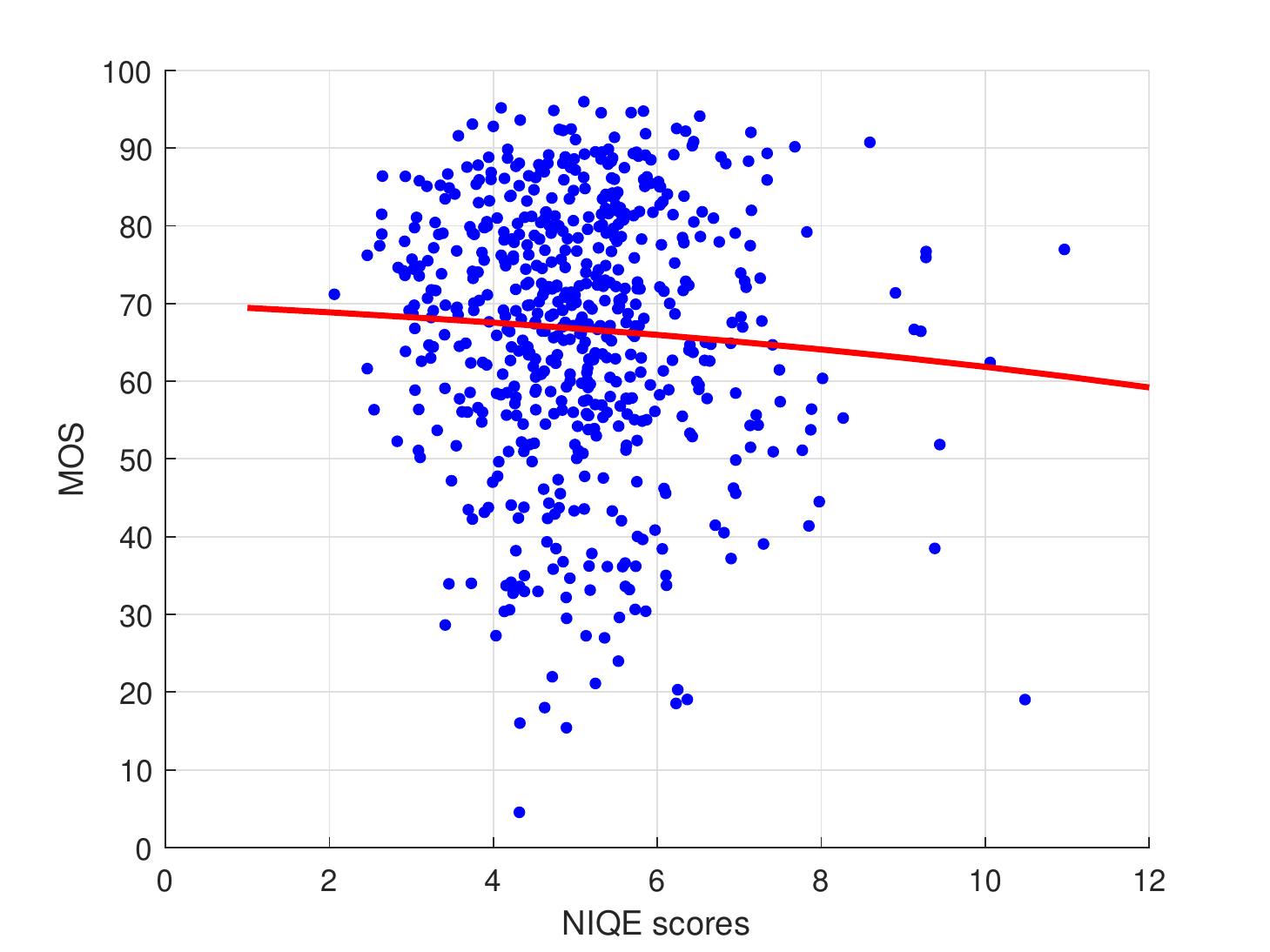}} 
 \subfigure[]{
   \label{performance_scatter:VGG16}
   \includegraphics[width=0.48\columnwidth]{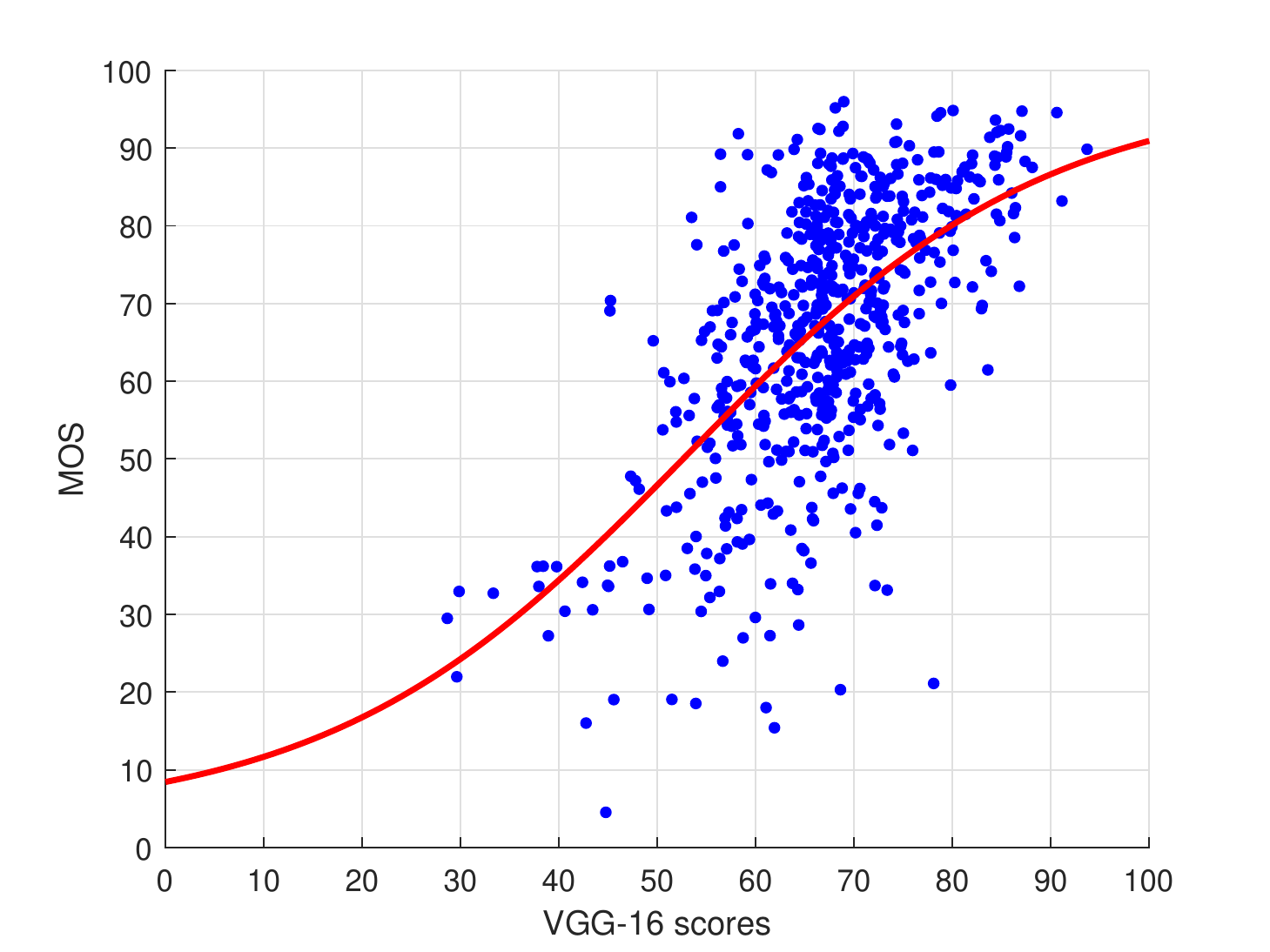}}    
 \subfigure[]{
   \label{performance_scatter:RAPIQUE}
   \includegraphics[width=0.48\columnwidth]{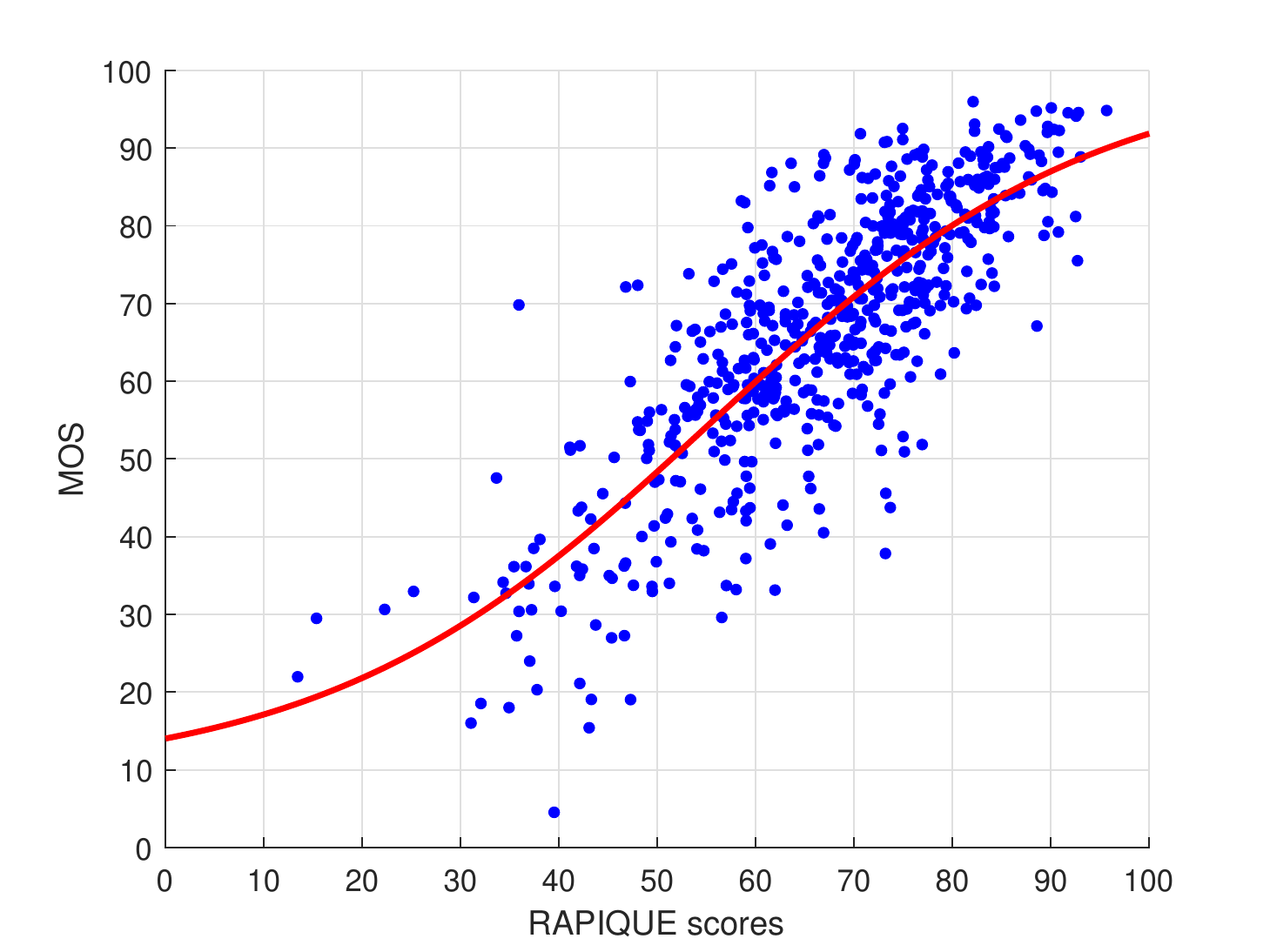}}
 \subfigure[]{
   \label{performance_scatter:NSS} 
   \includegraphics[width=0.48\columnwidth]{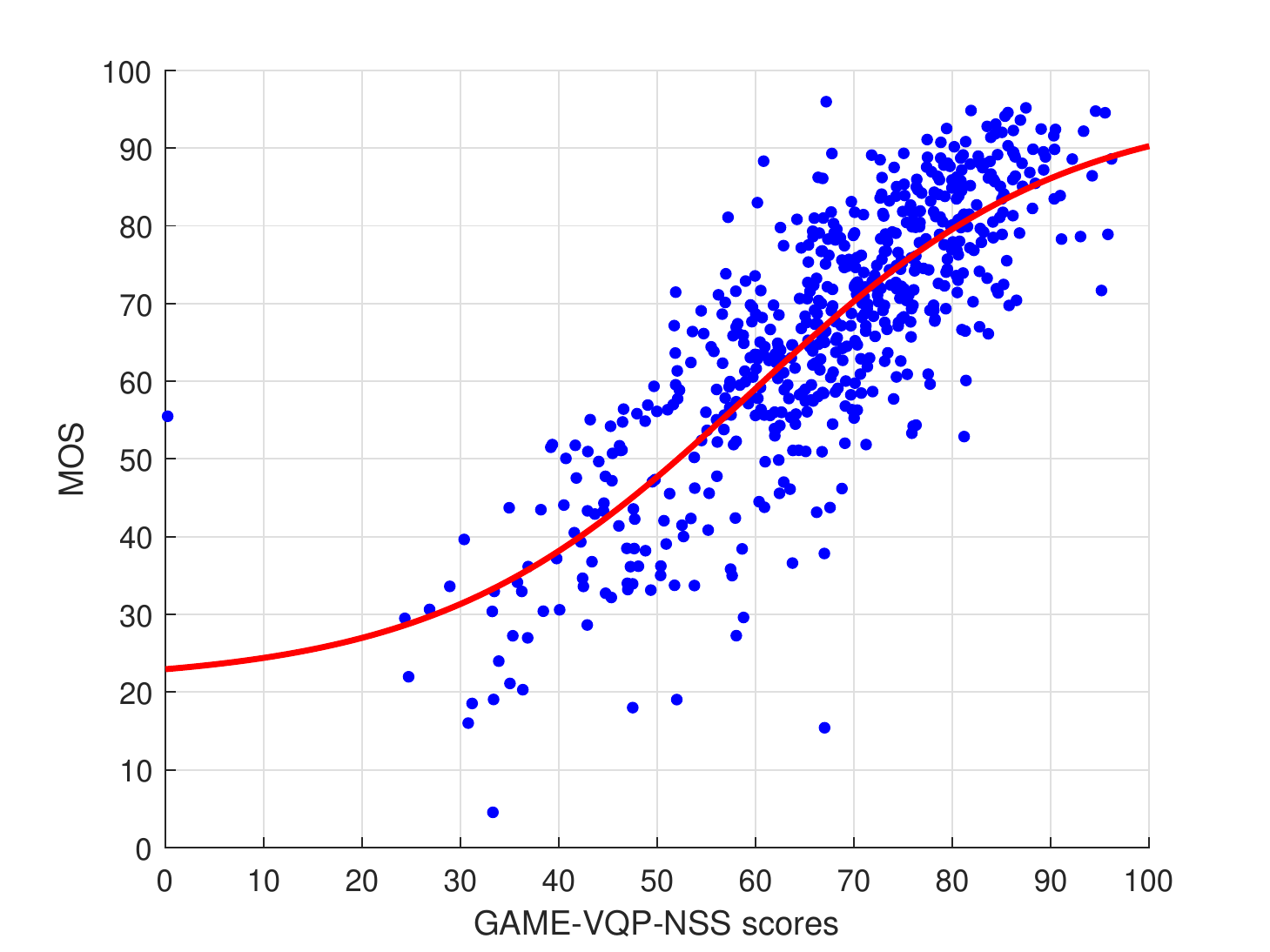}}
 \subfigure[]{
   \label{performance_scatter:CNN} 
   \includegraphics[width=0.48\columnwidth]{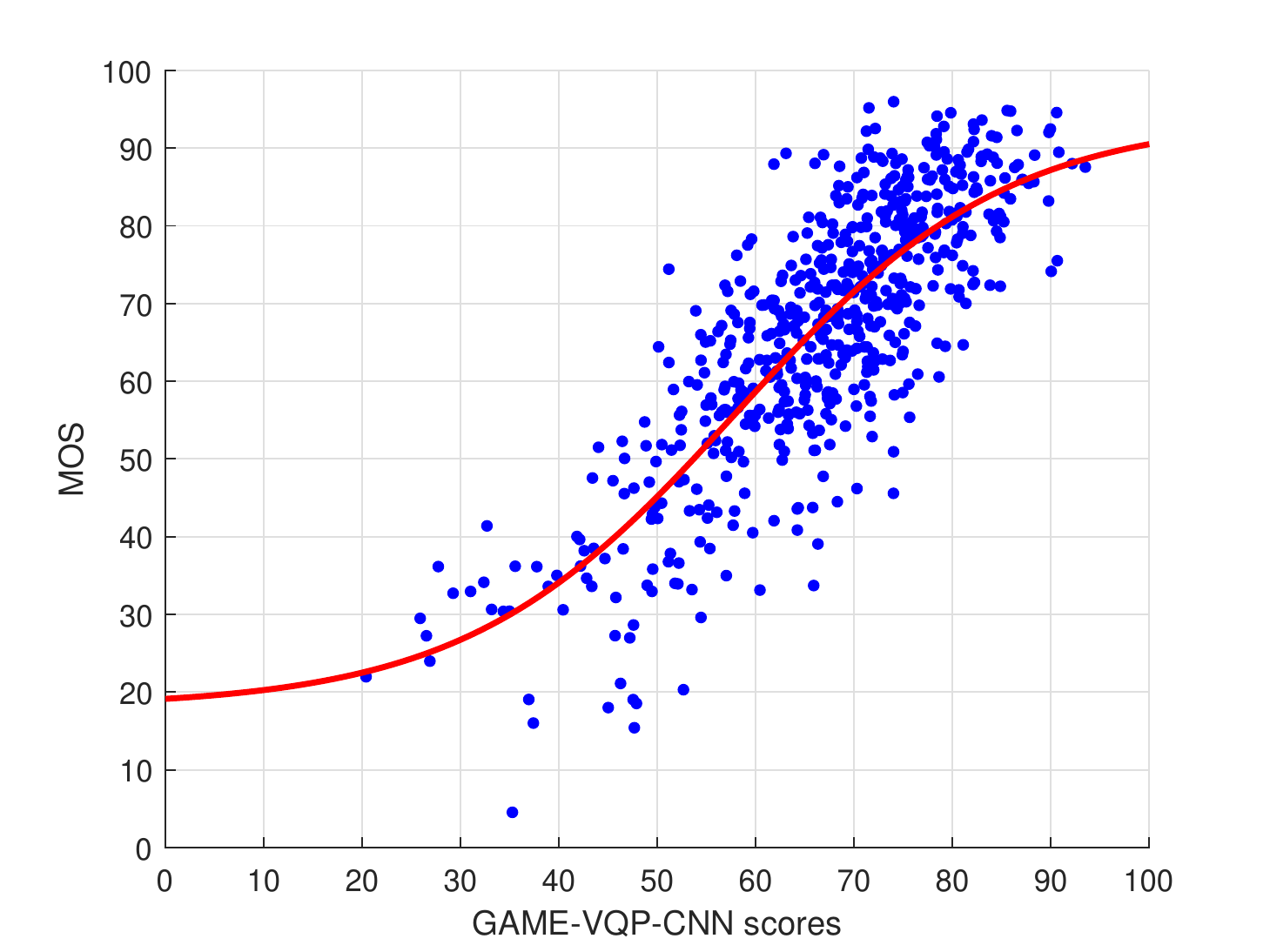}}
 \subfigure[]{
   \label{performance_scatter:GAME-VQP} 
   \includegraphics[width=0.48\columnwidth]{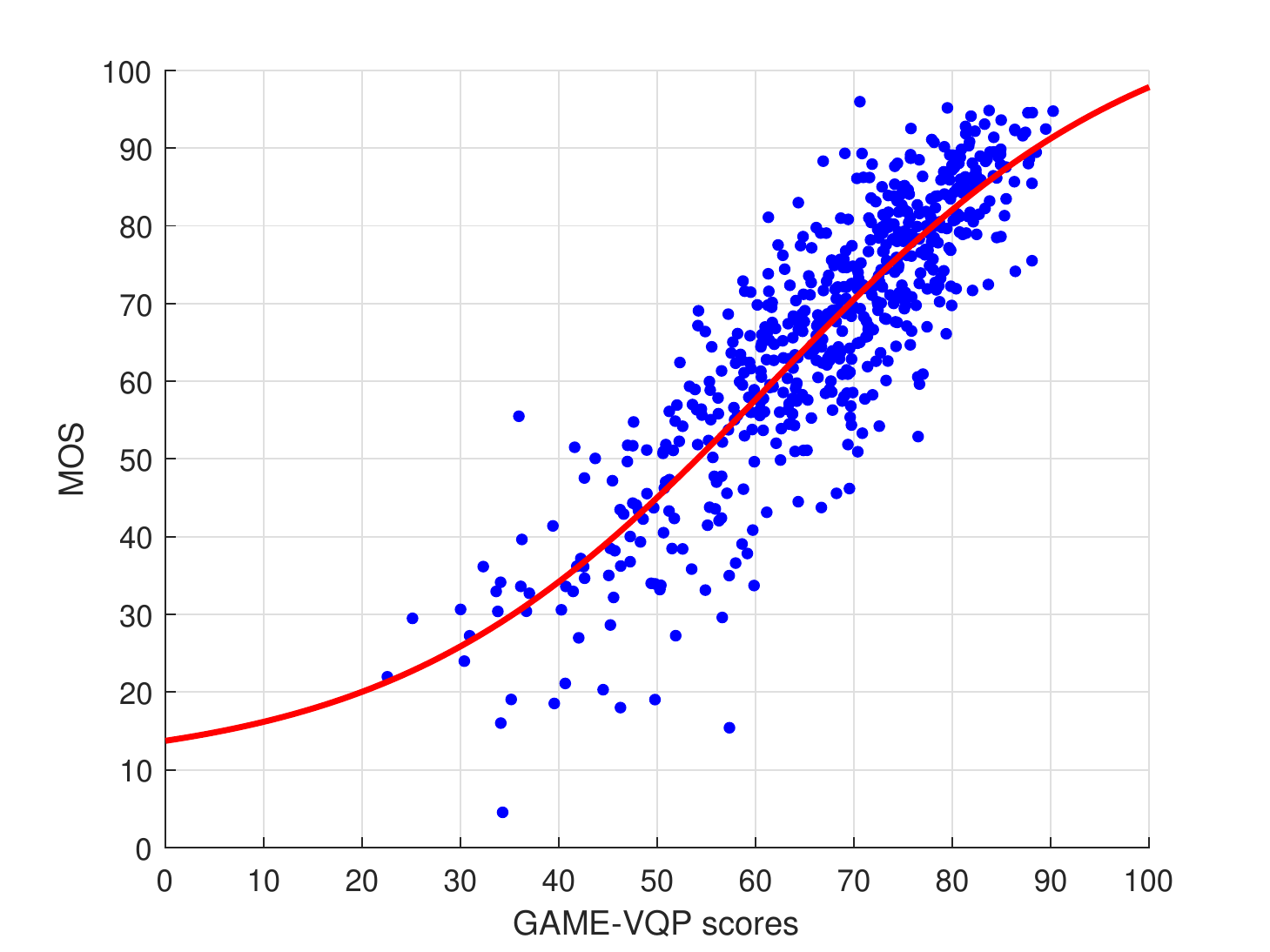}}
 \caption{Scatter plots of predicted quality scores versus MOS trained with an SVR using 5-fold cross validation on the LIVE-YouTube Gaming Video Quality Database. VQA models: (a) NIQE, (b) VGG-16 (c) RAPIQUE, (d) GAME-VQP-NSS (using NSS features only), (e) GAME-VQP-CNN (using CNN features only), (f) GAME-VQP. }
 \label{performance_scatter} 
\end{figure}

We applied 5-fold cross validation on several of the compared models, as well as the NSS and deep parts of GAME-VQP on the LIVE-YT-Gaming dataset, and aggregated the predicted scores obtained from each fold. 
Scatter plots of the quality predictions from NIQE, VGG-16, RAPIQUE, the NSS part of GAME-VQP, the CNN part of GAME-VQP, and the complete GAME-VQP against MOS are shown in Fig. \ref{performance_scatter}. 
Clearly, the distribution of GAME-VQP prediction scores is more compact than those of the other compared models/modules, indicating it has a stronger, more reliable correlation with MOS.

\section{Conclusion}
\label{conclusion}
This paper discusses research on an emerging application space, gaming video VQA, and proposes a new video quality prediction model called GAME-VQP that is specifically designed for UGC gaming videos. 
GAME-VQP deploys a variety of relevant NSS models on which parametric features are defined and used for quality prediction. 
A pre-trained CNN model is also used, whereby the output of the final average pooling layer is used to supply quality-relevant semantic features. 
Two independent SVR models are trained on the NSS and CNN features respectively, then the final predicted score is taken to be the mean of the two score predictions. 
Extensive testing on a large new gaming quality database, called LIVE-YT-Gaming, demonstrates the superior performance of the new model against existing SOTA algorithms. 
Using a standard UGC VQA database, we also showed that GAME-VQP is versatile enough to be used for the quality prediction of other types of UGC videos.

% or


%\appendices

%\section{}

\section*{Acknowledgment}

This research was supported by a gift from YouTube, and by grant number 2019844 from the National Science Foundation AI Institute for Foundations of Machine Learning (IFML).

\ifCLASSOPTIONcaptionsoff
  \newpage
\fi



%\bibliographystyle{IEEEtran}
% \end{thebibliography}

%%%% Reference %%%%
%%%%%%%%%%%%%%%%%%%%%%%%%%%%%%%%%
%%%%%%%%%%%%%%%%%%%%%%%%%%%%%%%%%
\bibliographystyle{IEEEtran}
\bibliography{0_main_journal}



%\vfill


\end{document}